\documentclass[10pt,a4paper]{article}\usepackage[]{graphicx}\usepackage[]{color}
\makeatletter
\def\maxwidth{ %
  \ifdim\Gin@nat@width>\linewidth
    \linewidth
  \else
    \Gin@nat@width
  \fi
}
\makeatother

\definecolor{fgcolor}{rgb}{0.345, 0.345, 0.345}
\newcommand{\hlnum}[1]{\textcolor[rgb]{0.686,0.059,0.569}{#1}}%
\newcommand{\hlstr}[1]{\textcolor[rgb]{0.192,0.494,0.8}{#1}}%
\newcommand{\hlcom}[1]{\textcolor[rgb]{0.678,0.584,0.686}{\textit{#1}}}%
\newcommand{\hlopt}[1]{\textcolor[rgb]{0,0,0}{#1}}%
\newcommand{\hlstd}[1]{\textcolor[rgb]{0.345,0.345,0.345}{#1}}%
\newcommand{\hlkwa}[1]{\textcolor[rgb]{0.161,0.373,0.58}{\textbf{#1}}}%
\newcommand{\hlkwb}[1]{\textcolor[rgb]{0.69,0.353,0.396}{#1}}%
\newcommand{\hlkwc}[1]{\textcolor[rgb]{0.333,0.667,0.333}{#1}}%
\newcommand{\hlkwd}[1]{\textcolor[rgb]{0.737,0.353,0.396}{\textbf{#1}}}%

\usepackage{framed}
\makeatletter
\newenvironment{kframe}{%
 \def\at@end@of@kframe{}%
 \ifinner\ifhmode%
  \def\at@end@of@kframe{\end{minipage}}%
  \begin{minipage}{\columnwidth}%
 \fi\fi%
 \def\FrameCommand##1{\hskip\@totalleftmargin \hskip-\fboxsep
 \colorbox{shadecolor}{##1}\hskip-\fboxsep
     \hskip-\linewidth \hskip-\@totalleftmargin \hskip\columnwidth}%
 \MakeFramed {\advance\hsize-\width
   \@totalleftmargin\z@ \linewidth\hsize
   \@setminipage}}%
 {\par\unskip\endMakeFramed%
 \at@end@of@kframe}
\makeatother

\definecolor{shadecolor}{rgb}{.97, .97, .97}
\definecolor{messagecolor}{rgb}{0, 0, 0}
\definecolor{warningcolor}{rgb}{1, 0, 1}
\definecolor{errorcolor}{rgb}{1, 0, 0}
\newenvironment{knitrout}{}{} 

\usepackage{alltt}
\usepackage[utf8x]{inputenc}
\usepackage{amsmath}
\usepackage{amsfonts}
\usepackage{amssymb}
\usepackage{graphicx}
\usepackage{url}
\usepackage{fancyvrb}
\usepackage[unicode=true,pdfusetitle,
 bookmarks=true,bookmarksnumbered=true,bookmarksopen=true,bookmarksopenlevel=2,
 breaklinks=false,pdfborder={0 0 1},backref=false,colorlinks=false]
 {hyperref}
\hypersetup{
 pdfstartview={XYZ null null 1}}
\usepackage{breakurl}

\newcommand{\PE}{package \texttt{performanceEstimation}\ }
\newcommand{\PEversion}{1.0.2\ }

\author{Luis Torgo\\FCUP - LIAAD/INESC Tec\\University of Porto\\
  \texttt{ltorgo@dcc.fc.up.pt}, \texttt{ltorgo@inesctec.pt}}
\title{An Infra-Structure for Performance Estimation\\ and Experimental Comparison\\ of Predictive Models in R}
\IfFileExists{upquote.sty}{\usepackage{upquote}}{}
\begin{document}

\maketitle

\begin{abstract}
  
  This document describes an infra-structure provided by the R \PE
  that allows to estimate the predictive performance of different
  approaches (workflows) to predictive tasks.  The infra-structure is
  generic in the sense that it can be used to estimate the values of
  any performance metrics, for any workflow on different predictive
  tasks, namely, classification, regression and time series tasks. The
  package also includes several standard workflows that allow users to
  easily set up their experiments limiting the amount of work and
  information they need to provide. The overall goal of the
  infra-structure provided by our package is to facilitate the task of
  estimating the predictive performance of different modeling
  approaches to predictive tasks in the R environment.
   
\end{abstract}

\section{Introduction}

The goal of this document is to describe the infra-structure that is
available in \PE\footnote{This document was written for version \PEversion of the package.} to estimate the performance of different approaches
to predictive tasks.  The main goal of this package is to provide a
general infra-structure that can be used to estimate the performance
using several predictive performance metrics of any modelling approach
to different predictive tasks, with a minimal effort from the
user. The package provides this type of facilities for classification,
regression and time series tasks. There is no limitation on the type
of approaches to these tasks for which you can estimate the
performance - the user just needs to provide a workflow function
(following some interface rules) that implements the approach for
which the predictive performance is to be estimated. The package
includes some standard workflow functions that implement the typical
learn+test approach that most users will be interested
in. This means that if you just want to estimate the performance of
some variants of a method already implemented in R (e.g. an SVM), on
some particular tasks, you will be able to use these standard workflow
functions and thus your required input will be limited to the minimum. The
package also provides a series of predictive performance metrics for
different tasks. Still, you are not limited to these metrics and can
use any metric as long as it exists in R or you provide a function
calculating it. Finally, the package also
include several standard data pre-processing and predictions
post-processing steps that again can be incorporated into the
workflows being evaluated.  

This infra-structure implements different methods for estimating the
predictive performance. Namely, you can select among: (i) cross
validation, (ii) holdout and random sub-sampling, (iii) leave one out
cross validation, (iv)  bootstrap ($\epsilon_0$ and .631) and also (v)
Monte-Carlo experiments for time series forecasting tasks. For each of
these tasks different options are implemented (e.g. use of stratified
sampling).

Most of the times experimental methodologies for performance estimation are iterative
processes that repeat the modeling task several times using different
train+test samples with the goal of improving the accuracy of the
estimates. The estimates are the result of aggregating the scores
obtained on each of the repetitions. For each of these
repetitions different training and testing samples are generated and
the process being evaluated is "asked" to: (i) obtain the predictive
model using the training data, and then (ii) use this model to obtain
predictions for the respective test sample. These predictions can then be
used to calculate the scores of the performance metrics being
estimated. This means that there is a workflow that starts with a
predictive task for which training and testing samples are given, and
that it should produce as result the predictions of the workflow for
the given test sample. There are far too many possible approaches and
sub-steps for the implementation of this workflow.  To ensure full
generality of the infra-structure, we allow the user to provide a
function that implements this workflow for each of the predictive
approaches she/he wishes to compare and/or evaluate. This function can
be parameterizable in the sense that there may be variants of the
workflow that the user wishes to evaluate and/or compare. Still, the
goal of this workflow  functions is very clear: (i)
receive as input a predictive task for which training  and  test
samples are given, as well as any eventual workflow specific parameters; and
(ii) produce as result a set of predictions for the given test
set.  These predictions
will then be used to obtain the scores of the predictive metrics for which the
user is interested in obtaining reliable estimates.

The infra-structure we describe here provides means for the user to
indicate: (i) a set of predictive tasks with the respective data sets;
(ii) a set of workflows and respective variants; and (iii) the
information on the estimation task. The infra-structure then takes care of all
the process of experimentally estimating the predictive performance of the different approaches on
the tasks, producing as result an object that can be
explored in different ways to obtain the results of the estimation process.
The infra-structure also provides several utility functions
to explore these results objects, for instance to
obtain summaries of the estimation process both in textual format as well as
visually. Moreover, it also provides functions that carry out
statistical significance tests based on the outcome of the
experiments. 

Finally, the infra-structure provides utility functions
implementing frequently used workflows for common modelling techniques, several data pre-processing steps and prediction post-processing operations, as
well as functions that facilitate the automatic generation of variants
of workflows by specifying sets of parameters that the user wishes to
consider in the comparisons.

\section{Package Installation}\label{sec:install}

The package can be installed as any other R package that is available at the R central repository (CRAN), i.e. by simply executing the code:

\begin{knitrout}\footnotesize
\definecolor{shadecolor}{rgb}{0.969, 0.969, 0.969}\color{fgcolor}\begin{kframe}
\begin{alltt}
\hlkwd{install.packages}\hlstd{(}\hlstr{"performanceEstimation"}\hlstd{)}
\end{alltt}
\end{kframe}
\end{knitrout}

This is the recommended form of installing the package an we assume that you start by doing this before you proceed to try the examples described in this document. This will install the current stable version of the package that at the time of writing of this document is version \PEversion.

If you wish to try some eventual new developments of the package that are still under testing and have not yet been released to the general public (thus are subject to bugs), then you may install the development version of the package that is available at the package GitHub Web page:

\url{https://github.com/ltorgo/performanceEstimation}

In order to install this development version (again not recommended unless you have a good reason for it), you may issue the following commands in R:

\begin{knitrout}\footnotesize
\definecolor{shadecolor}{rgb}{0.969, 0.969, 0.969}\color{fgcolor}\begin{kframe}
\begin{alltt}
\hlkwd{library}\hlstd{(devtools)}  \hlcom{# You need to install this package before!}
\hlkwd{install_github}\hlstd{(}\hlstr{"ltorgo/performanceEstimation"}\hlstd{,}\hlkwc{ref}\hlstd{=}\hlstr{"develop"}\hlstd{)}
\end{alltt}
\end{kframe}
\end{knitrout}

The GitHub Web page referred above is also the right place for you to report any issues you have with the package or help in its development.

\section{A Simple Illustrative Example}\label{sec:simpleEx}

Let us assume we are interested in estimating the predictive performance of several variants of
an SVM on the
\textbf{Iris} classification problem. More specifically, we want to
obtain a reliable estimate of the error rate of these variants using
10-fold cross validation. The following code illustrates how these
estimates could be obtained with our proposed infra-structure.

\begin{knitrout}\footnotesize
\definecolor{shadecolor}{rgb}{0.969, 0.969, 0.969}\color{fgcolor}\begin{kframe}
\begin{alltt}
\hlkwd{library}\hlstd{(performanceEstimation)}  \hlcom{# Loading our infra-structure}
\hlkwd{library}\hlstd{(e1071)}                  \hlcom{# A package containing SVMs}

\hlkwd{data}\hlstd{(iris)}                      \hlcom{# The data set we are going to use}

\hlstd{res} \hlkwb{<-} \hlkwd{performanceEstimation}\hlstd{(}
         \hlkwd{PredTask}\hlstd{(Species} \hlopt{~} \hlstd{.,iris),}
         \hlkwd{Workflow}\hlstd{(}\hlstr{"standardWF"}\hlstd{,}\hlkwc{learner}\hlstd{=}\hlstr{"svm"}\hlstd{),}
         \hlkwd{EstimationTask}\hlstd{(}\hlkwc{metrics}\hlstd{=}\hlstr{"err"}\hlstd{,}\hlkwc{method}\hlstd{=}\hlkwd{CV}\hlstd{())}
         \hlstd{)}
\end{alltt}
\begin{verbatim}
## 
## 
## ##### PERFORMANCE ESTIMATION USING  CROSS VALIDATION  #####
## 
## ** PREDICTIVE TASK :: iris.Species
## 
## ++ MODEL/WORKFLOW :: svm 
## Task for estimating  err  using
##  1 x 10 - Fold Cross Validation
## 	 Run with seed =  1234 
## Iteration :  1  2  3  4  5  6  7  8  9  10
\end{verbatim}
\end{kframe}
\end{knitrout}

This simple example illustrates several key concepts of our
infra-structure. First of all, we have the main function  -
\texttt{performanceEstimation()}, which is used to carry out
the estimation process. It has 3 main arguments: (i) a vector of
predictive tasks (in the example a single one); (ii) a vector of workflows (in the example a single one); and (iii) the
specification of the estimation task (essentially the metrics to be estimated and the estimation methodology). The package defines three classes of objects for storing the information related with these three concepts: predictive tasks (S4 class \textbf{PredTask}); workflows (S4 class \textbf{Workflow}); and estimation tasks (S4 class \textbf{EstimationTask}).

\textbf{PredTask} objects can be created by providing information on the formula defining the task, the source data set (an R data frame or the name of such object), and an optional ID (a string) to give to the task.

\textbf{Workflow} objects include information on the name of the function implementing the workflow and any number of parameters to be passed to this function when it will be called with different train and test sets.

\textbf{EstimationTask} objects provide information on the metrics for which we want a reliable estimate and also the methodology to be used to obtain these estimation. In case of metrics not defined with the \PE we also need to supply the name of the function that can be used to calculate these specific metrics.

In the above simple example we are using the data frame \textbf{iris} to create a task consisting of forecasting the variable \textit{Species} using all remaining variables. We are solving this classification task by using the function \texttt{standardWF} with the parameter \textbf{learner} set to ``svm''. This workflow function is already provided by the package and essentially can be used for the most frequent situations where a user just wants to apply and existing learning method to solve some task. This avoids the need for the user to have to write the functions solving the task. The \texttt{standardWF} function simply applies the modeling function indicated through the parameter \textbf{learner} (in this case the function \texttt{svm} defined in package \textbf{e1071}~\cite{e1071}) to the training set, and uses the resulting model to obtain the predictions for the test set. As we will see later the \texttt{standardWF} function also implements a series of common data pre-processing steps (e.g. filling in unknown values), and several common prediction post-processing steps (e.g. applying some transformation to the predicted values). Finally, the above example specifies the estimation task as using 10-fold Cross Validation to obtain estimates of the error rate.

The result of the call to \texttt{performanceEstimation()} is an S4
object of the class \textbf{ComparisonResults}. These objects tipically are not
 directly explored by the end-user so we ommit their
details here\footnote{Interested readers may have a look at the corresponding
  help page - \texttt{class?ComparisonResults} .}. There are several utility
functions that allow the users to explore the results of the
experimental comparisons, as shown by the next few illustrative examples.


\begin{knitrout}\footnotesize
\definecolor{shadecolor}{rgb}{0.969, 0.969, 0.969}\color{fgcolor}\begin{kframe}
\begin{alltt}
\hlkwd{summary}\hlstd{(res)}
\end{alltt}
\begin{verbatim}
## 
## == Summary of a  Cross Validation Performance Estimation Experiment ==
## 
## Task for estimating  err  using
##  1 x 10 - Fold Cross Validation
## 	 Run with seed =  1234 
## 
## * Predictive Tasks ::  iris.Species
## * Workflows  ::  svm 
## 
## -> Task:  iris.Species
##   *Workflow: svm 
##                err
## avg     0.03333333
## std     0.04714045
## med     0.00000000
## iqr     0.06666667
## min     0.00000000
## max     0.13333333
## invalid 0.00000000
\end{verbatim}
\end{kframe}
\end{knitrout}

The generic function \texttt{summary} allows us to obtain the
estimated scores for each compared approach on each predictive
task. For each performance metric (in this case only the error rate),
the function shows several descriptive statistics of the performance of the workflow in the different iterations of the estimation process. Moreover,
information is also given on eventual failures on some of the
iterations.



The generic function \texttt{plot} can be used to obtain a graphical
display of the distribution of performance metrics across the
different iterations of the estimation process using box-plots, as
show in Figure~\ref{fig:ex1Iris}. 

\begin{knitrout}\footnotesize
\definecolor{shadecolor}{rgb}{0.969, 0.969, 0.969}\color{fgcolor}\begin{kframe}
\begin{alltt}
\hlkwd{plot}\hlstd{(res)}
\end{alltt}
\end{kframe}\begin{figure}

{\centering \includegraphics[width=0.7\textwidth]{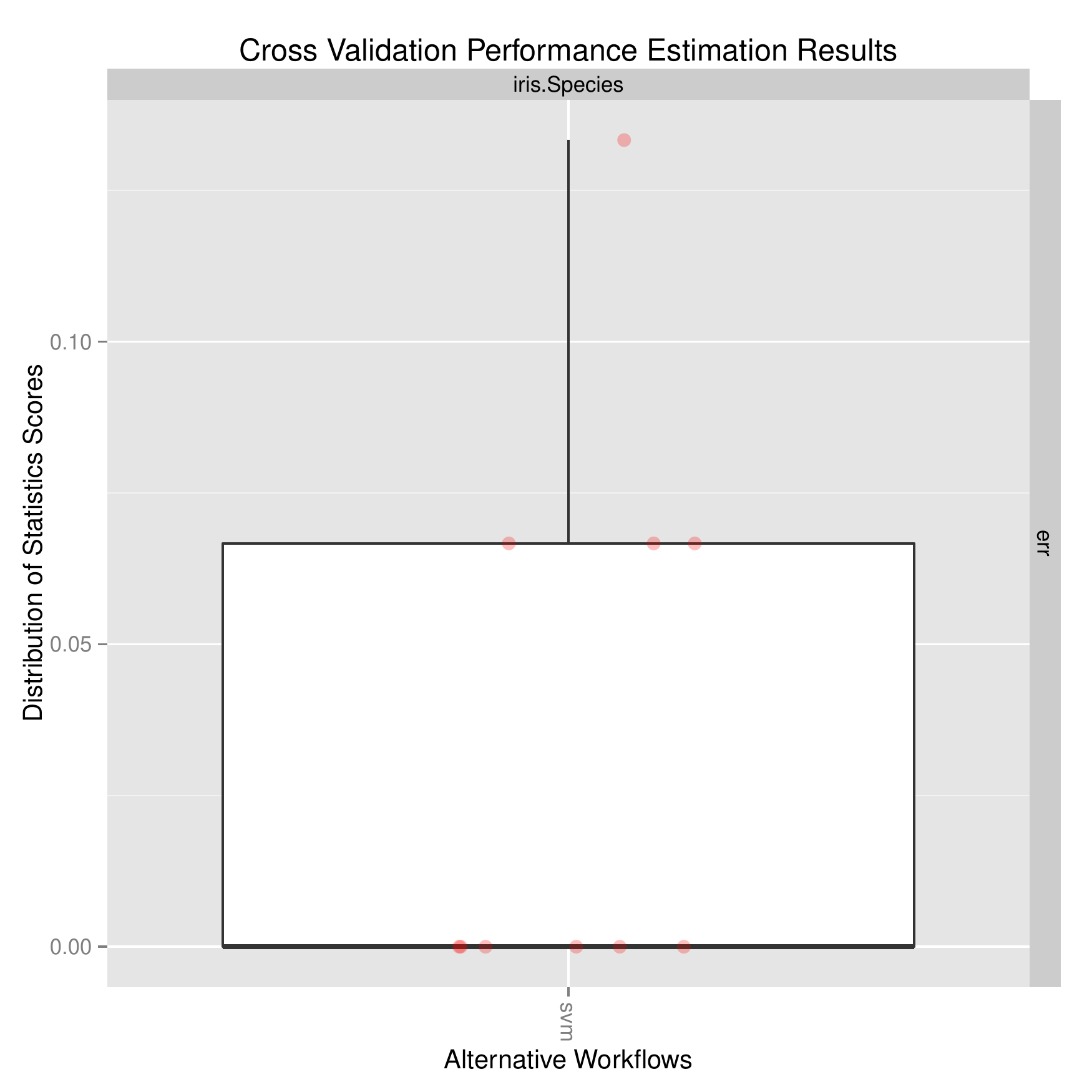} 

}

\caption[The distribution of the error rate on the 10 folds]{The distribution of the error rate on the 10 folds. Red dots show the performance on each individual iteration of the estimation process.}\label{fig:ex1Iris}
\end{figure}

\end{knitrout}



The above  example is just a simple illustration of the key concepts of the \PE. Most of the times you will use the package for more complex experiments, usually involving several tasks, several workflows or workflow variants, and eventually estimating several performance metrics. Before, we provide illustrations for these more complex setups we describe in more detail the key concepts of the \PE.

\section{Predictive Tasks}

Predictive tasks are data analysis problems where we want to obtain a
model of an unknown function $Y=f(X_1, X_2, \cdots, X_p)$ that relates
a target variable $Y$ with a set of $p$ predictors $X_1, X_2, \cdots,
X_p$. The model is usually obtained using a sample of $n$ observations
of the mapping of the unknown function, $D=\{\langle \mathbf{x}_i,
Y_i\rangle\}_{i=1}^n$, where $\mathbf{x}_i$ is a vector with the $p$
predictors values and $Y_i$ the respective target variable value.  These data sets in R are usually stored in data
frames, and formula objects are used to specify the form of the
functional dependency that we are trying to model, i.e. which is the
target variable and the predictors.

Objects of class \textbf{PredTask} encapsulate the information of a
predictive task, i.e. the functional form and the data required for
solving it. For convenience they also allow the user to assign a name
to each task. These S4 objects can be created using the construtor
function \texttt{PredTask()}, as seen in the following example:

\begin{knitrout}\footnotesize
\definecolor{shadecolor}{rgb}{0.969, 0.969, 0.969}\color{fgcolor}\begin{kframe}
\begin{alltt}
\hlkwd{data}\hlstd{(iris)}
\hlkwd{PredTask}\hlstd{(Species} \hlopt{~} \hlstd{.,iris)}
\end{alltt}
\begin{verbatim}
## Prediction Task Object:
## 	Task Name         :: iris.Species 
## 	Task Type         :: classification 
## 	Target Feature    :: Species 
## 	Formula           :: Species ~ .
## 	Task Data Source  :: iris
\end{verbatim}
\end{kframe}
\end{knitrout}

We should remark that, by default and for memory economy particularly if you are using large data sets, the objects of this class do not copy internally the data of the provided data frame. This  means that they assume that the data frame provided in the second argument will still exist when the estimation experiments will be executed. If you want the data to be copied into the \textbf{PredTask} object then you should call the constructor as follows:

\begin{knitrout}\footnotesize
\definecolor{shadecolor}{rgb}{0.969, 0.969, 0.969}\color{fgcolor}\begin{kframe}
\begin{alltt}
\hlkwd{data}\hlstd{(iris)}
\hlkwd{PredTask}\hlstd{(Species} \hlopt{~} \hlstd{.,iris,}\hlkwc{copy}\hlstd{=}\hlnum{TRUE}\hlstd{)}
\end{alltt}
\begin{verbatim}
## Prediction Task Object:
## 	Task Name         :: iris.Species 
## 	Task Type         :: classification 
## 	Target Feature    :: Species 
## 	Formula           :: Species ~ .
## 	Task Data Source  :: internal  150x5 data frame.
\end{verbatim}
\end{kframe}
\end{knitrout}

Note the difference on the task data source information to indicate that the data set is stored internally (i.e. a copy was made from the provided object). 

\section{Workflows and Workflow Variants}

Estimation methodologies work most of the times by re-sampling the
available data set $D$ in order to create different train and test
samples from $D$ (an exception being a single repetition of hold-out). The goal is to estimate the predictive performance
of a proposed workflow to solve the task, by using these different
samples to increase our confidence on the estimates. This workflow
consists on the process of obtaining a model from a given training
sample and then use it to obtain predictions for the given test
set. This process can include several steps, e.g. specific data
pre-processing steps, and may use any modelling approach, eventually
developed  by the user. 

We distinguish two types of workflows: (i) generic, and (ii) user-defined. The first are probably the most common setups where the user simply wants to apply methods that already exist in R to some task. The second covers situations where the user has developed her/his own specific approach to solve a task and wants to apply it to some concrete problems. The package covers both situations as we will see.

Independently of the workflows being generic or user-defined, they will typically have parameters that we can tune to try to improve the results. Frequently, we want to estimate and compare the performance of these parameter variants over some set of tasks. Our package also facilitates these setups by providing a function to easily specify these alternatives.

\subsection{Workflow Variants}\label{sec:variants}

The function \texttt{workflowVariants}  provides  means to easily create a vector of \textbf{Workflow} objects, each resulting from a variation of some parameter values of a certain workflow. For instance, suppose you want to evaluate the performance of SVMs on a certain task and you wish to consider a certain range of values for the parameters \textbf{cost} and \textbf{gamma} of the function \texttt{svm} from package \textbf{e1071}. You could obviously create a vector of \textbf{Workflow} objects, one for each combination of \textbf{cost} and \textbf{gamma} you wish to consider in your experiment. This could become very easily an annoying task if you want to consider a large number of combinations. Instead you can use the \texttt{workflowVariants} function.

The following example illustrates its use by considering 15 variants of SVMs applied to the Boston Housing data:

\begin{knitrout}\footnotesize
\definecolor{shadecolor}{rgb}{0.969, 0.969, 0.969}\color{fgcolor}\begin{kframe}
\begin{alltt}
\hlkwd{library}\hlstd{(performanceEstimation)}
\hlkwd{library}\hlstd{(e1071)}
\hlkwd{data}\hlstd{(Boston,}\hlkwc{package}\hlstd{=}\hlstr{"MASS"}\hlstd{)}

\hlstd{exp} \hlkwb{<-} \hlkwd{performanceEstimation}\hlstd{(}
         \hlkwd{PredTask}\hlstd{(medv} \hlopt{~} \hlstd{.,Boston),}
         \hlkwd{workflowVariants}\hlstd{(}\hlkwc{wf}\hlstd{=}\hlstr{"standardWF"}\hlstd{,}\hlkwc{learner}\hlstd{=}\hlstr{"svm"}\hlstd{,}
                          \hlkwc{learner.pars}\hlstd{=}\hlkwd{list}\hlstd{(}\hlkwc{cost}\hlstd{=}\hlnum{1}\hlopt{:}\hlnum{5}\hlstd{,}\hlkwc{gamma}\hlstd{=}\hlkwd{c}\hlstd{(}\hlnum{0.1}\hlstd{,}\hlnum{0.05}\hlstd{,}\hlnum{0.01}\hlstd{))),}
         \hlkwd{EstimationTask}\hlstd{(}\hlkwc{metrics}\hlstd{=}\hlstr{"mae"}\hlstd{,}\hlkwc{method}\hlstd{=}\hlkwd{CV}\hlstd{())}
         \hlstd{)}
\end{alltt}
\end{kframe}
\end{knitrout}

The function accepts any workflow function in parameter
\textbf{wf}. All remaining parameters are taken as parameters of that
workflow function. If you include  any of these parameters with a vector of values
(as it is the case of \textbf{cost} and \textbf{gamma} in the above
example), the values in these vectors will be used to generate
variants of the workflow. The variants are essentially all possible
combinations of the values in each vector. In the above example we
provide 5 values for \textbf{cost} and 3 for \textbf{gamma} and thus
we will have 15 \textbf{Workflow} objects. It may happen that some of
the parameters of the workflow actually take as valid values a
vector. In those situations you do not want the function to use the
values to generate variants. You can achieve this effect by including
the name of that parameter in the parameter \textbf{as.is} of the
function \texttt{workflowVariants}.

\subsection{Generic Workflows}

Most of the times users want to evaluate standard workflows on some tasks. This means that they want to use existing out-of-the-box tools within R and apply them to some data sets.  As such, 
the most frequently used workflows will essentially build a model
using some existing algorithm and obtain its predictions for the test set. To facilitate this type of approaches, \PE includes generic workflow functions. The idea is to save the user
from having to write her/his own workflow functions provided her/his workflow fits
this generic schema.

\subsubsection{Classification and Regression Tasks}\label{sec:standardWF}

Function \texttt{standardWF()} implements a typical workflow for both
classification and regression tasks. Apart from a formula, a training set data frame and a test set data frame, this function has the following main parameters that help the user to specify the intended approach:

\begin{description}
\item[learner] - the name of an R function that obtains a model from the training data. This function will be called with a formula in the first argument and the training set data frame in the second. This will typically be the name of some existing learning algorithm in R.
\item[learner.pars] - a list specifying any extra parameter settings that should be added to the formula and training set, at the time the learner function is called (defaults to \texttt{NULL}).
\item[predictor] - the name of an R function that is able to obtain the predictions of the model obtained with \texttt{learner}. This function will be called with the object resulting from the \texttt{learner} call on the first argument and the test set data frame in the second (it defaults to function \texttt{predict}). Most of the times you will not change this default because most modeling techniques in R have a predict method that can be applied to the models.
\item[predictor.pars] - a list specifying any extra parameter settings that should be added to the model and test set, at the time the predictor function is called (defaults to \texttt{NULL}). This is sometimes handy has some predict methods of some algorithms allow for instance to either return class labels or probabilities, and the option is made through some extra parameter when calling the predict function.
\end{description}

Below you find another simple example that illustrates the use of some of these parameters, this time to estimate the accuracy of 3 variants of a classification tree using 2$\times$5-CV.

\begin{knitrout}\footnotesize
\definecolor{shadecolor}{rgb}{0.969, 0.969, 0.969}\color{fgcolor}\begin{kframe}
\begin{alltt}
\hlkwd{library}\hlstd{(performanceEstimation)}
\hlkwd{library}\hlstd{(DMwR)}
\hlkwd{data}\hlstd{(iris)}

\hlstd{res} \hlkwb{<-} \hlkwd{performanceEstimation}\hlstd{(}
         \hlkwd{PredTask}\hlstd{(Species} \hlopt{~} \hlstd{.,iris),}
         \hlkwd{workflowVariants}\hlstd{(}\hlkwc{learner}\hlstd{=}\hlstr{"rpartXse"}\hlstd{,}
                          \hlkwc{learner.pars}\hlstd{=}\hlkwd{list}\hlstd{(}\hlkwc{se}\hlstd{=}\hlkwd{c}\hlstd{(}\hlnum{0}\hlstd{,}\hlnum{0.5}\hlstd{,}\hlnum{1}\hlstd{)),}
                          \hlkwc{predictor.pars}\hlstd{=}\hlkwd{list}\hlstd{(}\hlkwc{type}\hlstd{=}\hlstr{"class"}\hlstd{)),}
         \hlkwd{EstimationTask}\hlstd{(}\hlkwc{metrics}\hlstd{=}\hlstr{"acc"}\hlstd{,}\hlkwc{method}\hlstd{=}\hlkwd{CV}\hlstd{(}\hlkwc{nReps}\hlstd{=}\hlnum{2}\hlstd{,}\hlkwc{nFolds}\hlstd{=}\hlnum{5}\hlstd{))}
         \hlstd{)}
\end{alltt}
\end{kframe}
\end{knitrout}

Notice how we have omitted the specification of the workflow name in the call to \texttt{workflowVariants}. If you do not specify it through parameter \textbf{wf} the function will assume that you want to use the workflow defined by the function \texttt{standardWF}, unless there is a parameter \textbf{type} in the call which will lead to assume a time series generic workflow provided by function \texttt{timeseriesWF} that will be described in Section~\ref{sec:timeseriesWF}. In effect, you can also use the same speedup trick when calling directly the \textbf{Workflow} constructor, i.e. \\
\texttt{Workflow("standardWF",learner="svm")} \\
is equivalent to \\
\texttt{Workflow(learner="svm")}.

We now present an example of one of the most frequent type of
comparisons users carry out - checking which is the ``best'' model for
a given predictive task. Let us restrict the search to a small set of
models for illustrative purposes and let us play with the well-known
Boston housing regression task:

\begin{knitrout}\footnotesize
\definecolor{shadecolor}{rgb}{0.969, 0.969, 0.969}\color{fgcolor}\begin{kframe}
\begin{alltt}
\hlkwd{data}\hlstd{(Boston,}\hlkwc{package}\hlstd{=}\hlstr{'MASS'}\hlstd{)}
\hlkwd{library}\hlstd{(DMwR)}
\hlkwd{library}\hlstd{(e1071)}
\hlkwd{library}\hlstd{(randomForest)}
\hlstd{bostonRes} \hlkwb{<-} \hlkwd{performanceEstimation}\hlstd{(}
  \hlkwd{PredTask}\hlstd{(medv} \hlopt{~} \hlstd{.,Boston),}
  \hlkwd{workflowVariants}\hlstd{(}\hlkwc{learner}\hlstd{=}\hlkwd{c}\hlstd{(}\hlstr{'rpartXse'}\hlstd{,}\hlstr{'svm'}\hlstd{,}\hlstr{'randomForest'}\hlstd{)),}
  \hlkwd{EstimationTask}\hlstd{(}\hlkwc{metrics}\hlstd{=}\hlstr{"mse"}\hlstd{,}\hlkwc{method}\hlstd{=}\hlkwd{CV}\hlstd{())}
  \hlstd{)}
\end{alltt}
\end{kframe}
\end{knitrout}

Notice that on this simple example we have used all modeling tools
with their default parameter settings which is not necessarily a good
idea when we are looking for the best performance. Still, the goal of
this illustration is to show you how simple this type of experiments
can be if you are using a standard workflow setting. In case you want
to use the modelling tools with other parameter settings then you
should separate them in different \texttt{workflowVariants} calls because each learner will most probably use different parameters, as shown
in the following example:

\begin{knitrout}\footnotesize
\definecolor{shadecolor}{rgb}{0.969, 0.969, 0.969}\color{fgcolor}\begin{kframe}
\begin{alltt}
\hlkwd{data}\hlstd{(Boston,}\hlkwc{package}\hlstd{=}\hlstr{'MASS'}\hlstd{)}
\hlkwd{library}\hlstd{(DMwR)}
\hlkwd{library}\hlstd{(e1071)}
\hlkwd{library}\hlstd{(randomForest)}
\hlstd{bostonRes} \hlkwb{<-} \hlkwd{performanceEstimation}\hlstd{(}
  \hlkwd{PredTask}\hlstd{(medv} \hlopt{~} \hlstd{.,Boston),}
  \hlkwd{c}\hlstd{(}\hlkwd{workflowVariants}\hlstd{(}\hlkwc{learner}\hlstd{=}\hlstr{'rpartXse'}\hlstd{,}
                     \hlkwc{learner.pars}\hlstd{=}\hlkwd{list}\hlstd{(}\hlkwc{se}\hlstd{=}\hlkwd{c}\hlstd{(}\hlnum{0}\hlstd{,}\hlnum{1}\hlstd{))),}
    \hlkwd{workflowVariants}\hlstd{(}\hlkwc{learner}\hlstd{=}\hlstr{'svm'}\hlstd{,}
                     \hlkwc{learner.pars}\hlstd{=}\hlkwd{list}\hlstd{(}\hlkwc{cost}\hlstd{=}\hlkwd{c}\hlstd{(}\hlnum{1}\hlstd{,}\hlnum{5}\hlstd{),}\hlkwc{gamma}\hlstd{=}\hlkwd{c}\hlstd{(}\hlnum{0.01}\hlstd{,}\hlnum{0.1}\hlstd{))),}
    \hlkwd{workflowVariants}\hlstd{(}\hlkwc{learner}\hlstd{=}\hlstr{'randomForest'}\hlstd{,}
                     \hlkwc{learner.pars}\hlstd{=}\hlkwd{list}\hlstd{(}\hlkwc{ntree}\hlstd{=}\hlkwd{c}\hlstd{(}\hlnum{500}\hlstd{,}\hlnum{1000}\hlstd{)))),}
  \hlkwd{EstimationTask}\hlstd{(}\hlkwc{metrics}\hlstd{=}\hlstr{"mse"}\hlstd{,}\hlkwc{method}\hlstd{=}\hlkwd{CV}\hlstd{())}
  \hlstd{)}
\end{alltt}
\end{kframe}
\end{knitrout}

Notice that this code involves estimating the mean squared error on the Boston Housing task for 8 different models through
10-fold cross validation (the \texttt{CV} function assumes \texttt{nFolds=10} by default).

On top of using out-of-the-box existing algorithms users also frequently apply standard data pre-processing steps before the models are obtained. Our \texttt{standardWF} function also implements a few examples of these steps that you can include in your comparisons. The following is a list of the parameters of this function that control these steps:

\begin{description}
\item[pre] - A vector of function names that will be applied in sequence to the train
  and test data frames, generating new versions, i.e. a sequence of data
  pre-processing functions.
\item[pre.pars] - A named list of parameter values to be passed to the pre-processing functions.
\end{description}

We have implemented a few of these functions. Namely, in the \textbf{pre} argument you may use the following strings:

\begin{description}
\item[scale] - that scales (subtracts the mean and divides by the standard
  deviation) any numeric features on both the training and testing
  sets. Note that the mean and standard deviation are calculated using
  only the training sample.

\item[centralImp] - that fills in any \texttt{NA} values in both sets using
  the median value for numeric predictors and the mode for nominal
  predictors. Once again these centrality statistics are calculated
  using only the training set although they are applied to both train
  and test sets.

\item[na.omit] - that uses the R function \texttt{na.omit} to remove
  any rows containing \texttt{NA}'s  from both the training and test sets.

\item[undersampl] - this undersamples the training data cases that do not
  belong to the minority class (this pre-processing step is only
  available for classification tasks!). It takes the parameter
  \textbf{perc.under} that controls the level of undersampling
  (defaulting to 1, which means that there would be as many cases from
  the minority as from the other(s) class(es)).

\item[smote] - this operation uses the SMOTE~\cite{CBOK02}
  resampling algorithm to  generate a new training sample with a more
  ``balanced'' distribution of the target class (this pre-processing step
  is only available for classification tasks!). It takes the parameters
  \textbf{perc.under},  \textbf{perc.over} and \textbf{k} to control the
  algorithm. Read the  documentation of function \texttt{SMOTE} to
  know more details. 
\end{description}

Note that you can also write your own data pre-processing functions
provided you follow some protocol, and then use the name of your
functions in the \textbf{pre} argument. Check the help page of
function \texttt{standardPRE} to know the details on this simple
protocol to write your own pre-processing functions.

The following is a simple example of using data pre-processing steps within our provided generic workflows:

\begin{knitrout}\footnotesize
\definecolor{shadecolor}{rgb}{0.969, 0.969, 0.969}\color{fgcolor}\begin{kframe}
\begin{alltt}
\hlkwd{library}\hlstd{(performanceEstimation)}
\hlkwd{data}\hlstd{(algae,}\hlkwc{package}\hlstd{=}\hlstr{"DMwR"}\hlstd{)}

\hlstd{res} \hlkwb{<-} \hlkwd{performanceEstimation}\hlstd{(}
         \hlkwd{PredTask}\hlstd{(a1} \hlopt{~} \hlstd{.,algae[,}\hlnum{1}\hlopt{:}\hlnum{12}\hlstd{],}\hlstr{"AlgaA1"}\hlstd{),}
         \hlkwd{Workflow}\hlstd{(}\hlkwc{learner}\hlstd{=}\hlstr{"lm"}\hlstd{,}\hlkwc{pre}\hlstd{=}\hlkwd{c}\hlstd{(}\hlstr{"centralImp"}\hlstd{,}\hlstr{"scale"}\hlstd{)),}
         \hlkwd{EstimationTask}\hlstd{(}\hlkwc{metrics}\hlstd{=}\hlstr{"mae"}\hlstd{,}\hlkwc{method}\hlstd{=}\hlkwd{CV}\hlstd{())}
         \hlstd{)}
\end{alltt}
\begin{verbatim}
## 
## 
## ##### PERFORMANCE ESTIMATION USING  CROSS VALIDATION  #####
## 
## ** PREDICTIVE TASK :: AlgaA1
## 
## ++ MODEL/WORKFLOW :: lm 
## Task for estimating  mae  using
##  1 x 10 - Fold Cross Validation
## 	 Run with seed =  1234 
## Iteration :  1  2  3  4  5  6  7  8  9  10
\end{verbatim}
\end{kframe}
\end{knitrout}

Finally, standard workflows may also include some post-processing steps to be applied to the predictions of the model. These may include for instance some re-scaling of these predictions or even minimizing the risk of predictions through some cost-based approach. Function \texttt{standardWF} also accepts some parameters that control these post-processing steps. The following is a list of the parameters of this function that control these steps:

\begin{description}
\item[post] - A vector of function names that will be applied in sequence to the
  predictions of the model, generating a new version, i.e. a sequence of data
  post-processing functions.

\item[post.pars] - A named list of parameter values to be passed to the post-processing functions.
\end{description}

As with pre-processing steps you may also write your own prediction post-processing functions (check the help page of \texttt{standardPOST} for details). Still, we currently provide the following alternatives:

\begin{description}
\item[na2central] - this function fills in any \texttt{NA} predictions into
  either the median (numeric targets) or mode (nominal targets) of the
  target variable on the training set. Note that this is only applicable
  to predictions that are vectors of values.

\item[onlyPos] - in some numeric forecasting tasks the target variable
  takes only positive values. Nevertheless, some models may insist in
  forecasting negative values. This function casts these negative values
  to zero. Note that this is only applicable
  to predictions that are vectors of numeric values.

\item[cast2int] - in some numeric forecasting tasks the target variable
  takes only values within some interval. Nevertheless, some models may
  insist in forecasting  values outside of this interval. This function
  casts these values  into the nearest interval boundary. This function
  requires that you supply the limits of this interval through
  parameters \textbf{infLim} and \textbf{supLim}. Note that this is only
  applicable to predictions that are vectors of numeric values.

\item[maxutil] - maximize the utility of the predictions~\cite{Elk01}  of a
  classifier. This method is only applicable to classification tasks and
  to algorithms that are able to produce as predictions a vector of
  class probabilities for each test case, i.e. a matrix of probabilities
  for a given test set. The method requires a cost-benefit matrix to be
  provided through the parameter \textbf{cb.matrix}. For each test case,
  and given the probabilities estimated by the classifier and the cost
  benefit matrix, the method predicts the classifier that maximizes the
  utility of the prediction. This approach~\cite{Elk01} is a slight 'evolution' of
  the original idea~\cite{BFOS84} that only considered the costs of errors and not the
  benefits of the correct classifications as in the case of cost-benefit
  matrices we are using here. The parameter \textbf{cb.matrix} must
  contain a (square) matrix of dimension $NClasses\times NClasses$ where entry
  $X_{i,j}$ corresponds to the cost/benefit of predicting a test case as
  belonging to class $j$ when it is of class $i$. The diagonal of this
  matrix (correct predicitons) should contain positive numbers
  (benefits), whilst numbers outside of the matrix should contain
  negative numbers (costs of misclassifications). See the Examples
  section of the help page of the function \texttt{standardPOST} for an illustration. 
\end{description}

In the next example we illustrate the use of the post-processing routines by ``correcting'' the predictions of a linear regression model regards the frequency of occurrence of an alga, which can not be below zero:

\begin{knitrout}\footnotesize
\definecolor{shadecolor}{rgb}{0.969, 0.969, 0.969}\color{fgcolor}\begin{kframe}
\begin{alltt}
\hlkwd{library}\hlstd{(performanceEstimation)}
\hlkwd{data}\hlstd{(algae,}\hlkwc{package}\hlstd{=}\hlstr{"DMwR"}\hlstd{)}

\hlstd{res} \hlkwb{<-} \hlkwd{performanceEstimation}\hlstd{(}
         \hlkwd{PredTask}\hlstd{(a1} \hlopt{~} \hlstd{.,algae[,}\hlnum{1}\hlopt{:}\hlnum{12}\hlstd{],}\hlstr{"AlgaA1"}\hlstd{),}
         \hlkwd{c}\hlstd{(}\hlkwd{Workflow}\hlstd{(}\hlkwc{wfID}\hlstd{=}\hlstr{"lm"}\hlstd{,}
                    \hlkwc{learner}\hlstd{=}\hlstr{"lm"}\hlstd{,}
                    \hlkwc{pre}\hlstd{=}\hlkwd{c}\hlstd{(}\hlstr{"centralImp"}\hlstd{,}\hlstr{"scale"}\hlstd{)),}
           \hlkwd{Workflow}\hlstd{(}\hlkwc{wfID}\hlstd{=}\hlstr{"lmOnlyPos"}\hlstd{,}
                    \hlkwc{learner}\hlstd{=}\hlstr{"lm"}\hlstd{,}
                    \hlkwc{pre}\hlstd{=}\hlkwd{c}\hlstd{(}\hlstr{"centralImp"}\hlstd{,}\hlstr{"scale"}\hlstd{),}
                    \hlkwc{post}\hlstd{=}\hlkwd{c}\hlstd{(}\hlstr{"onlyPos"}\hlstd{))),}
         \hlkwd{EstimationTask}\hlstd{(}\hlkwc{metrics}\hlstd{=}\hlstr{"mae"}\hlstd{,}\hlkwc{method}\hlstd{=}\hlkwd{CV}\hlstd{())}
         \hlstd{)}
\end{alltt}
\begin{verbatim}
## 
## 
## ##### PERFORMANCE ESTIMATION USING  CROSS VALIDATION  #####
## 
## ** PREDICTIVE TASK :: AlgaA1
## 
## ++ MODEL/WORKFLOW :: lm 
## Task for estimating  mae  using
##  1 x 10 - Fold Cross Validation
## 	 Run with seed =  1234 
## Iteration :  1  2  3  4  5  6  7  8  9  10
## 
## ++ MODEL/WORKFLOW :: lmOnlyPos 
## Task for estimating  mae  using
##  1 x 10 - Fold Cross Validation
## 	 Run with seed =  1234 
## Iteration :  1  2  3  4  5  6  7  8  9  10
\end{verbatim}
\begin{alltt}
\hlkwd{summary}\hlstd{(res)}
\end{alltt}
\begin{verbatim}
## 
## == Summary of a  Cross Validation Performance Estimation Experiment ==
## 
## Task for estimating  mae  using
##  1 x 10 - Fold Cross Validation
## 	 Run with seed =  1234 
## 
## * Predictive Tasks ::  AlgaA1
## * Workflows  ::  lm, lmOnlyPos 
## 
## -> Task:  AlgaA1
##   *Workflow: lm 
##               mae
## avg     14.308536
## std      1.857755
## med     14.340022
## iqr      2.447854
## min     11.652404
## max     17.451525
## invalid  0.000000
## 
##   *Workflow: lmOnlyPos 
##               mae
## avg     13.116271
## std      1.690161
## med     13.173095
## iqr      2.273219
## min      9.955684
## max     15.302897
## invalid  0.000000
\end{verbatim}
\end{kframe}
\end{knitrout}

As you see this simple post-processing step improved the performance of the model considerably.

\subsubsection{Time Series Tasks}\label{sec:timeseriesWF}

Our infra-structure also includes another generic workflow function
that is specific for predictive tasks with time-dependent data
(e.g. time series forecasting problems). This workflow function
implements two different approaches to the problem of training a
 model with a set of time-dependent data and then use it to
obtain predictions for a test set in the future. These two approaches
contrast with the standard approach of learning a model with the
available training sample and then use it to obtain predictions for
all test period. This standard approach could be applied using the previously
described \texttt{standardWF()} function. However, there are
alternatives to this procedure, two of the most common being the
sliding and growing window approaches, which are implemented in another workflow function developed specifically for time series tasks.

Predictive tasks for time-dependent data are different from standard
classification and regression tasks because they require that the test
samples have time stamps that are more recent then training
samples. In this context, experimental methodologies handling these
tasks should not shuffle the observations to maintain the time ordering
of the original data. The most common setup is that we have a $L$  time steps
training window containing observations in the period $[t_1,t_L]$ and a $F$
time steps test window typically containing the observations in the
time window $[t_{L+1},t_{L+F}]$. 

The idea of the
sliding window method is that if we want a prediction for time point
$t_k$ belonging to the test interval $[t_{L+1},t_{L+F}]$ then we can
assume that all data from $t_{L+1}$ till $t_{k-1}$ is already past,
and thus usable by the model. In this context, it may be wise to use
this new data in the interval $[t_{L+1},t_{k-1}]$ to update the
original model obtained using only the initial training period data. This is
particularly advisable if we suspect that the conditions may have
changed since the training period has ended. Model updating using the
sliding window method is carried out by using the data in the $L$ last
time steps, i.e. every new model is always obtained using the last $L$
data observations, as if the training window was slided forward in
time. Our \texttt{timeseriesWF()} function implements this idea for
both time series with a numeric target variable and a nominal target
variable. This function has a parameter (\texttt{type}) that if set to
``slide'' will use a sliding window approach. As with the
\texttt{standardWF()} function, this \texttt{timeseriesWF()} function
also accepts parameters specifying the learner, predictor, 
and their respective parameters, as well as the previously defined pre- and post-processing steps, all with exactly the same names and possible values (c.f. Section ~\ref{sec:standardWF}). Moreover, this function also includes
an extra parameter, named \texttt{relearn.step}, which allows the user
to establish the frequency of model updating. By default this is every
new test sample, i.e. $1$, but the user may set a less frequent
model-updating policy by using higher values of this parameter to avoid high computation costs. 

The
idea of the growing window approach is very similar. The only difference
is on the data used when updating the models. Whilst sliding window
uses the data occurring in the last $L$ time steps, growing window
keeps increasing the original training window with the newly available
data points, i.e. the models are obtained with increasingly larger
training samples. By setting the parameter \texttt{type} to ``grow''
you get the \texttt{timseriesWF()} function to use this method.

The following code illustrates these two approaches by comparing them to the standard approach of using a single model to forecast all testing period:

\begin{knitrout}\footnotesize
\definecolor{shadecolor}{rgb}{0.969, 0.969, 0.969}\color{fgcolor}\begin{kframe}
\begin{alltt}
\hlkwd{library}\hlstd{(performanceEstimation)}
\hlkwd{library}\hlstd{(quantmod)}
\hlkwd{library}\hlstd{(randomForest)}
\hlkwd{getSymbols}\hlstd{(}\hlstr{'^GSPC'}\hlstd{,}\hlkwc{from}\hlstd{=}\hlstr{'2008-01-01'}\hlstd{,}\hlkwc{to}\hlstd{=}\hlstr{'2012-12-31'}\hlstd{)}
\hlstd{data.model} \hlkwb{<-} \hlkwd{specifyModel}\hlstd{(}
  \hlkwd{Next}\hlstd{(}\hlnum{100}\hlopt{*}\hlkwd{Delt}\hlstd{(}\hlkwd{Ad}\hlstd{(GSPC)))} \hlopt{~} \hlkwd{Delt}\hlstd{(}\hlkwd{Ad}\hlstd{(GSPC),}\hlkwc{k}\hlstd{=}\hlnum{1}\hlopt{:}\hlnum{10}\hlstd{))}
\hlstd{data} \hlkwb{<-} \hlkwd{as.data.frame}\hlstd{(}\hlkwd{modelData}\hlstd{(data.model))}
\hlkwd{colnames}\hlstd{(data)[}\hlnum{1}\hlstd{]} \hlkwb{<-} \hlstr{'PercVarClose'}
\hlstd{spExp} \hlkwb{<-} \hlkwd{performanceEstimation}\hlstd{(}
  \hlkwd{PredTask}\hlstd{(PercVarClose} \hlopt{~} \hlstd{.,data,}\hlstr{'SP500_2012'}\hlstd{),}
  \hlkwd{c}\hlstd{(}\hlkwd{Workflow}\hlstd{(}\hlkwc{wf}\hlstd{=}\hlstr{'standardWF'}\hlstd{,}\hlkwc{wfID}\hlstd{=}\hlstr{"standRF"}\hlstd{,}
             \hlkwc{learner}\hlstd{=}\hlstr{'randomForest'}\hlstd{,}
             \hlkwc{learner.pars}\hlstd{=}\hlkwd{list}\hlstd{(}\hlkwc{ntree}\hlstd{=}\hlnum{500}\hlstd{)),}
    \hlkwd{Workflow}\hlstd{(}\hlkwc{wf}\hlstd{=}\hlstr{'timeseriesWF'}\hlstd{,}\hlkwc{wfID}\hlstd{=}\hlstr{"slideRF"}\hlstd{,}
             \hlkwc{learner}\hlstd{=}\hlstr{'randomForest'}\hlstd{,}
             \hlkwc{learner.pars}\hlstd{=}\hlkwd{list}\hlstd{(}\hlkwc{ntree}\hlstd{=}\hlnum{500}\hlstd{),}
             \hlkwc{type}\hlstd{=}\hlstr{"slide"}\hlstd{,}
             \hlkwc{relearn.step}\hlstd{=}\hlnum{30}\hlstd{),}
    \hlkwd{Workflow}\hlstd{(}\hlkwc{wf}\hlstd{=}\hlstr{'timeseriesWF'}\hlstd{,}\hlkwc{wfID}\hlstd{=}\hlstr{"growRF"}\hlstd{,}
             \hlkwc{learner}\hlstd{=}\hlstr{'randomForest'}\hlstd{,}
             \hlkwc{learner.pars}\hlstd{=}\hlkwd{list}\hlstd{(}\hlkwc{ntree}\hlstd{=}\hlnum{500}\hlstd{),}
             \hlkwc{type}\hlstd{=}\hlstr{"grow"}\hlstd{,}
             \hlkwc{relearn.step}\hlstd{=}\hlnum{30}\hlstd{)}
   \hlstd{),}
  \hlkwd{EstimationTask}\hlstd{(}\hlkwc{metrics}\hlstd{=}\hlkwd{c}\hlstd{(}\hlstr{"mse"}\hlstd{,}\hlstr{"theil"}\hlstd{),}
                 \hlkwc{method}\hlstd{=}\hlkwd{MonteCarlo}\hlstd{(}\hlkwc{nReps}\hlstd{=}\hlnum{5}\hlstd{,}\hlkwc{szTrain}\hlstd{=}\hlnum{0.5}\hlstd{,}\hlkwc{szTest}\hlstd{=}\hlnum{0.25}\hlstd{))}
  \hlstd{)}
\end{alltt}
\end{kframe}
\end{knitrout}

The above example applies 3 different workflows to the task of trying to forecast the percentage daily variation of the prices of S\&P 500, using some information of the previous prices as predictors. Namely, all workflows use a random forest with 500 trees but the predictions for each test set of the 5 repetitions Monte Carlo estimation methodology (c.f. Section~\ref{sec:MC}), are obtained differently. The first workflow, named ``standRF'', obtains a single random forest with the training set and uses it to obtain predictions for the full test set. The other two approaches, instead of using this standard workflow, take advantage of the workflow provided by \texttt{timeseriesWF} and use either  sliding  or growing windows to obtain these predictions. For both these two latter approaches a new random forest is obtained after each 30 new test cases (set by parameter \texttt{relearn.step}).

\subsection{User-defined Workflows}

With the goal of ensuring that the
proposed infra-structure is able to cope with all  possible usage
scenarios, we also allow the user to write and provide her/his own workflow functions to be used in the estimation tasks, provided they follow some protocol. Namely, these
user-defined workflow functions should be written such that the
first three parameters are: (i) the formula defining the predictive
task; (ii) the provided training sample; and (iii) the test sample
for which predictions are to be obtained. The functions may eventually
accept other arguments with specific parameters of the user-defined workflow. The
following is a general sketch of a user-defined workflow function:

\begin{knitrout}\footnotesize
\definecolor{shadecolor}{rgb}{0.969, 0.969, 0.969}\color{fgcolor}\begin{kframe}
\begin{alltt}
\hlstd{myWorkFlow} \hlkwb{<-} \hlkwa{function}\hlstd{(}\hlkwc{form}\hlstd{,}\hlkwc{train}\hlstd{,}\hlkwc{test}\hlstd{,}\hlkwc{...}\hlstd{) \{}
  \hlkwd{require}\hlstd{(mySpecialPackage,}\hlkwc{quietly}\hlstd{=}\hlnum{TRUE}\hlstd{)}
  \hlcom{## cary out some data pre-processing}
  \hlstd{myTrain} \hlkwb{<-} \hlkwd{mySpecificPreProcessingSteps}\hlstd{(train)}
  \hlcom{## now obtain the model}
  \hlstd{myModel} \hlkwb{<-} \hlkwd{myModelingTechnique}\hlstd{(form,myTrain,...)}
  \hlcom{## obtain the predictions}
  \hlstd{preds} \hlkwb{<-} \hlkwd{predict}\hlstd{(myModel,test)}
  \hlcom{## cary out some predictions post-processing}
  \hlstd{newPreds} \hlkwb{<-} \hlkwd{mySpecificPostProcessingSteps}\hlstd{(form,train,test,preds)}
  \hlkwd{names}\hlstd{(newPreds)} \hlkwb{<-} \hlkwd{rownames}\hlstd{(test)}
  \hlcom{## finally produce the list containing the output of the workflow}
  \hlstd{res} \hlkwb{<-} \hlkwd{list}\hlstd{(}\hlkwc{trues}\hlstd{=}\hlkwd{responseValues}\hlstd{(form,test),}\hlkwc{preds}\hlstd{=newPreds)}
  \hlkwd{return}\hlstd{(res)}
\hlstd{\}}
\end{alltt}
\end{kframe}
\end{knitrout}

Not all workflows will require all these steps, though some may even
require more. This is clearly something that is up to the user. The
only strict requirements for these functions are: (i) the first 3
parameters of the workflow function should be the formula, train and
test data frames; and (ii) the result of the function should be a list. 

Regarding the components of this results list, in case you wish to use the functions of the package that calculate some standard prediction metrics, then you should make sure that this list contains at least a component named \texttt{trues} with the vector of the true values of the target variable in the test set, and another component named \texttt{preds} with the respective predictions produced by the workflow for this test cases.
 These lists returned by the workflows  may optionally contain any other information the creator of the workflow function deems
important to return. 

The sketch shown above also illustrates the use of the function
\texttt{responseValues()} that can be used to obtain the values of the target
variable given a formula and a data frame.

On top of the 3 mandatory parameters (formula, training and test sets), user-defined workflow functions may also accept any other arguments. As we have seen in Section~\ref{sec:variants} we provide
the function \texttt{workflowVariants()} to facilitate the specification of
different variants of any workflow function by trying all combinations
of several of its specific parameters. For instance, if the modelling
function in the above example workflow (function
\texttt{myModelingTechnique()}) had an integer parameter \texttt{x}
and a Boolean parameter \texttt{y}, we could generate several
\textbf{Workflow} objects to be evaluated/compared using the
\texttt{performanceEstimation()} function, as follows:

\begin{knitrout}\footnotesize
\definecolor{shadecolor}{rgb}{0.969, 0.969, 0.969}\color{fgcolor}\begin{kframe}
\begin{alltt}
\hlkwd{workflowVariants}\hlstd{(}\hlstr{'myWorkFlow'}\hlstd{,}\hlkwc{x}\hlstd{=}\hlkwd{c}\hlstd{(}\hlnum{0}\hlstd{,}\hlnum{3}\hlstd{,}\hlnum{5}\hlstd{,}\hlnum{7}\hlstd{),}\hlkwc{y}\hlstd{=}\hlkwd{c}\hlstd{(}\hlnum{TRUE}\hlstd{,}\hlnum{FALSE}\hlstd{))}
\end{alltt}
\end{kframe}
\end{knitrout}

This would generate 8 variants of the same workflow with all
combinations of the specified values for the 2 parameters.  

Let us see a concrete example of a user supplied workflow
function. Imagine we want to evaluate a kind of ensemble model formed
by a regression tree and a multiple linear regression model on an
algae blooms data set~\cite{Tor10}.  We write a workflow function that implements our intended workflow:

\begin{knitrout}\footnotesize
\definecolor{shadecolor}{rgb}{0.969, 0.969, 0.969}\color{fgcolor}\begin{kframe}
\begin{alltt}
\hlstd{RLensemble} \hlkwb{<-} \hlkwa{function}\hlstd{(}\hlkwc{f}\hlstd{,} \hlkwc{tr}\hlstd{,} \hlkwc{ts}\hlstd{,} \hlkwc{weightRT}\hlstd{=}\hlnum{0.5}\hlstd{,} \hlkwc{step}\hlstd{=}\hlnum{FALSE}\hlstd{,} \hlkwc{...}\hlstd{,} \hlkwc{.models}\hlstd{=}\hlnum{FALSE}\hlstd{) \{}
  \hlkwd{require}\hlstd{(DMwR,}\hlkwc{quietly}\hlstd{=}\hlnum{TRUE}\hlstd{)}
  \hlcom{## Getting the column id of the target variable}
  \hlstd{tgtCol} \hlkwb{<-} \hlkwd{which}\hlstd{(}\hlkwd{colnames}\hlstd{(tr)} \hlopt{==} \hlkwd{as.character}\hlstd{(f[[}\hlnum{2}\hlstd{]]))}
  \hlcom{## filling in NAs using knnImputation}
  \hlstd{noNAsTR} \hlkwb{<-} \hlstd{tr}
  \hlstd{noNAsTS} \hlkwb{<-} \hlstd{ts}
  \hlstd{noNAsTR[,}\hlopt{-}\hlstd{tgtCol]} \hlkwb{<-} \hlkwd{knnImputation}\hlstd{(tr[,}\hlopt{-}\hlstd{tgtCol])}
  \hlstd{noNAsTS[,}\hlopt{-}\hlstd{tgtCol]} \hlkwb{<-} \hlkwd{knnImputation}\hlstd{(ts[,}\hlopt{-}\hlstd{tgtCol],}\hlkwc{distData}\hlstd{=tr[,}\hlopt{-}\hlstd{tgtCol])}
  \hlstd{r} \hlkwb{<-} \hlkwd{rpartXse}\hlstd{(f,tr,...)}
  \hlstd{l} \hlkwb{<-} \hlkwd{lm}\hlstd{(f,noNAsTR)}
  \hlkwa{if} \hlstd{(step) l} \hlkwb{<-} \hlkwd{step}\hlstd{(l,}\hlkwc{trace}\hlstd{=}\hlnum{0}\hlstd{)}
  \hlstd{pr} \hlkwb{<-} \hlkwd{predict}\hlstd{(r,ts)}
  \hlstd{pl} \hlkwb{<-} \hlkwd{predict}\hlstd{(l,noNAsTS)}
  \hlstd{ps} \hlkwb{<-} \hlstd{weightRT}\hlopt{*}\hlstd{pr}\hlopt{+}\hlstd{(}\hlnum{1}\hlopt{-}\hlstd{weightRT)}\hlopt{*}\hlstd{pl}
  \hlkwd{names}\hlstd{(ps)} \hlkwb{<-} \hlkwd{rownames}\hlstd{(ts)}
  \hlstd{res} \hlkwb{<-} \hlkwd{list}\hlstd{(}\hlkwc{trues}\hlstd{=}\hlkwd{responseValues}\hlstd{(f,ts),}\hlkwc{preds}\hlstd{=ps)}
  \hlkwa{if} \hlstd{(.models) res} \hlkwb{<-} \hlkwd{c}\hlstd{(res,}\hlkwd{list}\hlstd{(}\hlkwc{linearModel}\hlstd{=l,}\hlkwc{tree}\hlstd{=r))}
  \hlstd{res}
\hlstd{\}}
\end{alltt}
\end{kframe}
\end{knitrout}

This workflow starts by building two modified samples of the training
and testing sets, with the \texttt{NA} values being filled in using a
nearest neighbour strategy (see the help page of the function
\texttt{knnImputation()} for more
details). These versions are to be used by the \texttt{lm()} function
that is unable to cope with cases with missing values. After obtaining
the two models and their predictions the function calculates a
weighted average of both predictions. Note that if \texttt{.models=TRUE} we add the models to the results list.

To evaluate different variants of this workflow we could run the
following experiment:

\begin{knitrout}\footnotesize
\definecolor{shadecolor}{rgb}{0.969, 0.969, 0.969}\color{fgcolor}\begin{kframe}
\begin{alltt}
\hlkwd{data}\hlstd{(algae,}\hlkwc{package}\hlstd{=}\hlstr{'DMwR'}\hlstd{)}
\hlstd{expRes} \hlkwb{<-} \hlkwd{performanceEstimation}\hlstd{(}
    \hlkwd{PredTask}\hlstd{(a1} \hlopt{~} \hlstd{.,algae[,}\hlnum{1}\hlopt{:}\hlnum{12}\hlstd{],}\hlstr{'alga1'}\hlstd{),}
    \hlkwd{workflowVariants}\hlstd{(}\hlstr{'RLensemble'}\hlstd{,}
                     \hlkwc{se}\hlstd{=}\hlkwd{c}\hlstd{(}\hlnum{0}\hlstd{,}\hlnum{1}\hlstd{),}\hlkwc{step}\hlstd{=}\hlkwd{c}\hlstd{(}\hlnum{TRUE}\hlstd{,}\hlnum{FALSE}\hlstd{),}\hlkwc{weightRT}\hlstd{=}\hlkwd{c}\hlstd{(}\hlnum{0.4}\hlstd{,}\hlnum{0.5}\hlstd{,}\hlnum{0.6}\hlstd{)),}
    \hlkwd{EstimationTask}\hlstd{(}\hlstr{"mse"}\hlstd{,}\hlkwc{method}\hlstd{=}\hlkwd{CV}\hlstd{()))}
\end{alltt}
\end{kframe}
\end{knitrout}

\section{Estimation Tasks}

The third argument of the main function \texttt{performanceEstimation} defines the estimation task we want to carry out. Its value is an S4 object of class \textbf{EstimationTask} that can be created through the constructor function with the same name. The main arguments of this constructor are \textbf{metrics} and \textbf{method}. The first is a vector with names of metrics for which we want reliable estimates, while the second is the method to be used to obtain these estimates. Other arguments of the constructor are \textbf{evaluator} and \textbf{evaluator.pars} that allow the specification of functions for calculating user-defined evaluation metrics. Finally, the last parameter of the constructor is \textbf{trainReq} that is a Boolean value indicating whether the training data should also be ``sent'' to the evaluation functions. This is useful for metrics that require the training data for being calculated (e.g. some normalized metrics use the average value of the target variable in the training set).

\subsection{Performance Metrics}

The package implements a reasonable set of the most common performance metrics. These include classification, regression and time series metrics. These are internally calculated by the functions \texttt{classificationMetrics} and \texttt{regressionMetrics}. The help pages of these two functions include an exhaustive list of these metrics together with their definitions. Depending on the predictive tasks being used one of these functions will be called to calculate the metrics the user indicates in the parameter \textbf{metrics} of the \texttt{EstimationTask} constructor. The user may also omit the \textbf{metrics} parameter and in that case all metrics available in the evaluation function will be calculated. On top of the metrics implemented in these functions the user may also include the strings ``trTime'', ``tsTime'' and ``totTime'' to obtain estimates of the training, testing and total computation times taken by the workflow. 


\subsection{User-defined Performance Metrics}

Although the functions \texttt{classificationMetrics} and \texttt{regressionMetrics} implement a large set of performance metrics, it is inevitable that some applications may require some domain-specific metrics for which users want reliable estimates. Our package allows the indication of user-defined functions that calculate such domain-specific metrics. This is achieved through parameters \textbf{evaluator} and \textbf{evaluator.pars} of the \textbf{EstimationTask} constructor. The first allows the user to specify the name of such function, whilst the second is a list with parameter values to use when calling such function. In order to be usable by our package such functions need to obey some input/output protocol. In terms of input parameters to these functions, our infra-structure will call these user-defined functions with: (i) the outcome of the workflow; (ii) any metric names that were specified in parameter \textbf{metrics}; (iii) the target values in the training set if the parameter \textbf{trainReq} was set to \texttt{TRUE}; and (iv) any other parameters specified through \textbf{evaluator.pars}. In terms of output, the user-defined functions must return as a result a (named) vector with as many scores as the number of metrics specified by the user. The names of the scores in the vector should be the names of the metrics specified in the parameter \textbf{metrics}.

Suppose we want to calculate the error of a regression model as the difference between true and predicted values raised to some power. We could define a function to calculate such metric following our package input/output protocol as:

\begin{knitrout}\footnotesize
\definecolor{shadecolor}{rgb}{0.969, 0.969, 0.969}\color{fgcolor}\begin{kframe}
\begin{alltt}
\hlstd{powErr} \hlkwb{<-} \hlkwa{function}\hlstd{(}\hlkwc{trues}\hlstd{,}\hlkwc{preds}\hlstd{,}\hlkwc{pow}\hlstd{=}\hlnum{3}\hlstd{,}\hlkwc{...}\hlstd{) \{}
    \hlkwd{c}\hlstd{(}\hlkwc{pow.err} \hlstd{=} \hlkwd{mean}\hlstd{((trues}\hlopt{-}\hlstd{preds)}\hlopt{^}\hlstd{pow))}
\hlstd{\}}
\end{alltt}
\end{kframe}
\end{knitrout}

Using such function in the context of some 10-fold cross validation estimation experiment would involve creating an estimation task as:

\begin{knitrout}\footnotesize
\definecolor{shadecolor}{rgb}{0.969, 0.969, 0.969}\color{fgcolor}\begin{kframe}
\begin{alltt}
\hlkwd{EstimationTask}\hlstd{(}\hlkwc{metrics}\hlstd{=}\hlstr{"pow.err"}\hlstd{,}\hlkwc{method}\hlstd{=}\hlkwd{CV}\hlstd{(),}
               \hlkwc{evaluator}\hlstd{=}\hlstr{"powErr"}\hlstd{,}\hlkwc{evaluator.pars}\hlstd{=}\hlkwd{list}\hlstd{(}\hlkwc{pow}\hlstd{=}\hlnum{4}\hlstd{))}
\end{alltt}
\end{kframe}
\end{knitrout}

In effect, you could take advantage of the defaults and actually just call it like this:

\begin{knitrout}\footnotesize
\definecolor{shadecolor}{rgb}{0.969, 0.969, 0.969}\color{fgcolor}\begin{kframe}
\begin{alltt}
\hlkwd{EstimationTask}\hlstd{(}\hlkwc{evaluator}\hlstd{=}\hlstr{"powErr"}\hlstd{,}\hlkwc{evaluator.pars}\hlstd{=}\hlkwd{list}\hlstd{(}\hlkwc{pow}\hlstd{=}\hlnum{4}\hlstd{))}
\end{alltt}
\end{kframe}
\end{knitrout}

The default of \textbf{method} is exactly Cross validation and the default of \textbf{metrics} is to calculate all metrics of the provided evaluator function.

\subsection{Estimation Methodologies}\label{sec:expMeth}

There are different ways of providing reliable estimates of the
predictive performance of a workflow. Our infra-structure implements some
of the most common estimation methods. In this section we
briefly describe them and provide short illustrative examples of their
use.

The parameter \textbf{method} of the \textbf{EstimationTask} constructor allows the user to specify the estimation methodology that will be used. The next sections explain the options available.

\subsubsection{Cross Validation}

$k$-Fold cross validation (CV) is one of the most common 
methods to estimate the predictive performance of a model. By
including an S4 object of class \textbf{CV} in the parameter \textbf{method} we can carry
out experiments of this type.

The constructor function \texttt{CV()} can be used to obtain objects
of class \textbf{CV}. It accepts the following parameters:

\begin{description}
\item[nReps] - the number of repetitions of the $k$-fold CV experiment (default is $1$)
\item[nFolds] - the number of $k$ folds to use (default is $10$)
\item[seed] - the random number generator seed to use (default is $1234$)
\item[strat] - whether to use stratified samples on the different iterations (default is \texttt{FALSE})
\item[dataSplits] - a list containing the data splits to
          use on each repetition of a k-folds CV experiment (defaulting
          to ‘NULL’). Check the help page of the
  class \textbf{CV} for further details. 
\end{description}

Bellow you can find a small illustration using the Breast Cancer data
set available in package \textbf{mlbench}. On this example we compare
some variants of an SVM using a $3\times 10-$fold cross validation
process with stratified sampling because one of the two classes has a
considerably lower frequency.

\begin{knitrout}\footnotesize
\definecolor{shadecolor}{rgb}{0.969, 0.969, 0.969}\color{fgcolor}\begin{kframe}
\begin{alltt}
\hlkwd{data}\hlstd{(BreastCancer,}\hlkwc{package}\hlstd{=}\hlstr{'mlbench'}\hlstd{)}
\hlkwd{library}\hlstd{(e1071)}
\hlstd{bcExp} \hlkwb{<-} \hlkwd{performanceEstimation}\hlstd{(}
  \hlkwd{PredTask}\hlstd{(Class} \hlopt{~} \hlstd{.,BreastCancer[,}\hlopt{-}\hlnum{1}\hlstd{],}\hlstr{'BreastCancer'}\hlstd{),}
  \hlkwd{workflowVariants}\hlstd{(}\hlstr{'standardWF'}\hlstd{,}
           \hlkwc{learner}\hlstd{=}\hlstr{'svm'}\hlstd{,}
           \hlkwc{learner.pars}\hlstd{=}\hlkwd{list}\hlstd{(}\hlkwc{cost}\hlstd{=}\hlkwd{c}\hlstd{(}\hlnum{1}\hlstd{,}\hlnum{5}\hlstd{),}\hlkwc{gamma}\hlstd{=}\hlkwd{c}\hlstd{(}\hlnum{0.01}\hlstd{,}\hlnum{0.1}\hlstd{))}
          \hlstd{),}
  \hlkwd{EstimationTask}\hlstd{(}\hlkwc{metrics}\hlstd{=}\hlkwd{c}\hlstd{(}\hlstr{"F"}\hlstd{,}\hlstr{"prec"}\hlstd{,}\hlstr{"rec"}\hlstd{),}
                 \hlkwc{evaluator.pars}\hlstd{=}\hlkwd{list}\hlstd{(}\hlkwc{posClass}\hlstd{=}\hlstr{"malignant"}\hlstd{),}
                 \hlkwc{method}\hlstd{=}\hlkwd{CV}\hlstd{(}\hlkwc{nReps}\hlstd{=}\hlnum{3}\hlstd{,}\hlkwc{nFolds}\hlstd{=}\hlnum{10}\hlstd{,}\hlkwc{strat}\hlstd{=}\hlnum{TRUE}\hlstd{)))}
\end{alltt}
\end{kframe}
\end{knitrout}

Please note the use of the \texttt{evaluator.pars} parameter of the
\texttt{EstimationTask} constructor function. We have used it to indicate which of the class labels of this problem should be considered the ``positive'' class, which is required to compute the values of the F, recall and precision metrics we are estimating through $3\times 10-$fold cross validation. This parameter setting is passed to the \texttt{classificationMetrics} function that is the default for classification tasks like Breast Cancer.

\subsubsection{Bootstrapping}

Bootstrapping or bootstrap resampling is another well-known
experimental methodology that is implemented in our
package. Namely, we implement two of the most common methods of obtaining bootstrap estimates: $\epsilon_0$ and $.632$ bootstrap.
By including an S4 object of class
\textbf{Bootstrap} in the \textbf{method} argument we can carry out experiments of this
type.

Function \texttt{Bootstrap()} can be used as a constructor of
objects of class \textbf{Bootstrap}. It accepts the following
arguments:

\begin{description}
\item[type] - a string with the type of bootstrap estimates: either "e0" for $\epsilon_0$ bootstrap, or ".632" for $.632$ bootstrap (default is "e0")
\item[nReps] - the number of repetitions of the bootstrap experiment (default is $200$)
\item[seed] - the random number generator seed to use (default is $1234$)
\item[dataSplits] - a list containing user-supplied data splits
  for each of the repetitions (check the help page of the
  class for further details). This parameter defaults to
  \texttt{NULL}, i.e. no user-supplied splits, they are decided
  internally by the infra-structure.
\end{description}

Bellow you can find a small illustration using the Servo data set available in package \textbf{mlbench}. On this example we compare some variants of an artificial neural network using 100 repetitions of a bootstrap experiment. 

\begin{knitrout}\footnotesize
\definecolor{shadecolor}{rgb}{0.969, 0.969, 0.969}\color{fgcolor}\begin{kframe}
\begin{alltt}
\hlkwd{data}\hlstd{(Servo,}\hlkwc{package}\hlstd{=}\hlstr{'mlbench'}\hlstd{)}
\hlkwd{library}\hlstd{(nnet)}
\hlstd{nnExp} \hlkwb{<-} \hlkwd{performanceEstimation}\hlstd{(}
  \hlkwd{PredTask}\hlstd{(Class} \hlopt{~} \hlstd{.,Servo),}
  \hlkwd{workflowVariants}\hlstd{(}\hlkwc{learner}\hlstd{=}\hlstr{'nnet'}\hlstd{,}
                   \hlkwc{learner.pars}\hlstd{=}\hlkwd{list}\hlstd{(}\hlkwc{trace}\hlstd{=F,}\hlkwc{linout}\hlstd{=T,}
                       \hlkwc{size}\hlstd{=}\hlkwd{c}\hlstd{(}\hlnum{3}\hlstd{,}\hlnum{5}\hlstd{),}\hlkwc{decay}\hlstd{=}\hlkwd{c}\hlstd{(}\hlnum{0.01}\hlstd{,}\hlnum{0.1}\hlstd{))}
          \hlstd{),}
  \hlkwd{EstimationTask}\hlstd{(}\hlkwc{metrics}\hlstd{=}\hlstr{"mse"}\hlstd{,}\hlkwc{method}\hlstd{=}\hlkwd{Bootstrap}\hlstd{(}\hlkwc{nReps}\hlstd{=}\hlnum{100}\hlstd{)))}
\end{alltt}
\end{kframe}
\end{knitrout}

\subsubsection{Holdout and Random Sub-sampling}

The Holdout is another frequently used experimental methodology,
particularly for large data sets. To carry out this type of
experiments in our infra-structure we can include an S4 object of
class \textbf{Holdout} in the third argument of function
\texttt{performanceEstimation()}.

Function \texttt{Holdout()} can be used as a constructor of
objects of class \textbf{Holdout}. It accepts the following
arguments:

\begin{description}
\item[nReps] - the number of repetitions of the Holdout experiment (default is $1$)
\item[hldSz] - the percentage  of cases (a number between 0 and 1) to leave as holdout (test set) (default is $0.3$)
\item[seed] - the random number generator seed to use (default is $1234$)
\item[strat] - whether to use stratified samples (default is \texttt{FALSE})
\item[dataSplits] - a list containing user-supplied data splits
  for each of the repetitions (check the help page of the
  class for further details). This parameter defaults to
  \texttt{NULL}, i.e. no user-supplied splits, they are decided
  internally by the infra-structure.
\end{description}

Please note that for the usual meaning of Holdout the number of repetitions should be 1 (the default), while larger values of this parameter correspond to what is usually known as random subsampling.

The following is a small illustrative example of the use of the
random subsampling with the LetterRecognition classification task from package
\textbf{mlbench}.

\begin{knitrout}\footnotesize
\definecolor{shadecolor}{rgb}{0.969, 0.969, 0.969}\color{fgcolor}\begin{kframe}
\begin{alltt}
\hlkwd{data}\hlstd{(LetterRecognition,}\hlkwc{package}\hlstd{=}\hlstr{'mlbench'}\hlstd{)}
\hlstd{ltrExp} \hlkwb{<-} \hlkwd{performanceEstimation}\hlstd{(}
    \hlkwd{PredTask}\hlstd{(lettr} \hlopt{~} \hlstd{.,LetterRecognition),}
    \hlkwd{workflowVariants}\hlstd{(}\hlkwc{learner}\hlstd{=}\hlstr{'rpartXse'}\hlstd{,}
                     \hlkwc{learner.pars}\hlstd{=}\hlkwd{list}\hlstd{(}\hlkwc{se}\hlstd{=}\hlkwd{c}\hlstd{(}\hlnum{0}\hlstd{,}\hlnum{1}\hlstd{)),}
                     \hlkwc{predictor.pars}\hlstd{=}\hlkwd{list}\hlstd{(}\hlkwc{type}\hlstd{=}\hlstr{'class'}\hlstd{)}
                     \hlstd{),}
    \hlkwd{EstimationTask}\hlstd{(}\hlkwc{metrics}\hlstd{=}\hlstr{"err"}\hlstd{,}\hlkwc{method}\hlstd{=}\hlkwd{Holdout}\hlstd{(}\hlkwc{nReps}\hlstd{=}\hlnum{3}\hlstd{,}\hlkwc{hldSz}\hlstd{=}\hlnum{0.3}\hlstd{)))}
\end{alltt}
\end{kframe}
\end{knitrout}

Please note the use of the \texttt{predictor.pars} parameter of our
\texttt{standardWF()} function to be able to cope with the fact that
the \texttt{predict} method for classification trees requires the use
of \texttt{type="class"} to get actual predicted class labels instead
of class probabilities.

\subsubsection{Leave One Out Cross Validation}

Leave one out cross validation is a type of cross validation method
that is mostly used for small data sets. You can think of leave one
out cross validation as a $k$-fold cross validation with $k$ equal to
the size of the available data set. To carry out this type of
experiments in our infra-structure we can include an S4 object of
class \textbf{LOCV} in the third argument of function
\texttt{performanceEstimation()}.

Function \texttt{LOOCV()} can be used as a constructor of
objects of class \textbf{LOOCV}. It accepts the following
arguments:

\begin{description}
\item[seed] - the random number generator seed to use (default is $1234$)
\item[dataSplits] - a list containing user-supplied data splits
  for each of the repetitions (check the help page of the
  class for further details). This parameter defaults to
  \texttt{NULL}, i.e. no user-supplied splits, they are decided
  internally by the infra-structure.
\end{description}

The following is a small illustrative example of the use of the
LOOCV with the Iris classification task.

\begin{knitrout}\footnotesize
\definecolor{shadecolor}{rgb}{0.969, 0.969, 0.969}\color{fgcolor}\begin{kframe}
\begin{alltt}
\hlkwd{data}\hlstd{(iris)}
\hlkwd{library}\hlstd{(e1071)}
\hlstd{irisExp} \hlkwb{<-} \hlkwd{performanceEstimation}\hlstd{(}
    \hlkwd{PredTask}\hlstd{(Species} \hlopt{~} \hlstd{.,iris),}
    \hlkwd{workflowVariants}\hlstd{(}\hlkwc{learner}\hlstd{=}\hlstr{'svm'}\hlstd{,}
                     \hlkwc{learner.pars}\hlstd{=}\hlkwd{list}\hlstd{(}\hlkwc{cost}\hlstd{=}\hlkwd{c}\hlstd{(}\hlnum{1}\hlstd{,}\hlnum{10}\hlstd{))}
                     \hlstd{),}
    \hlkwd{EstimationTask}\hlstd{(}\hlkwc{metrics}\hlstd{=}\hlstr{"acc"}\hlstd{,}\hlkwc{method}\hlstd{=}\hlkwd{LOOCV}\hlstd{()))}
\end{alltt}
\end{kframe}
\end{knitrout}

\subsubsection{Monte Carlo Experiments}\label{sec:MC}

\begin{figure}[ht]
  \centering
  \includegraphics[width=0.65\textwidth]{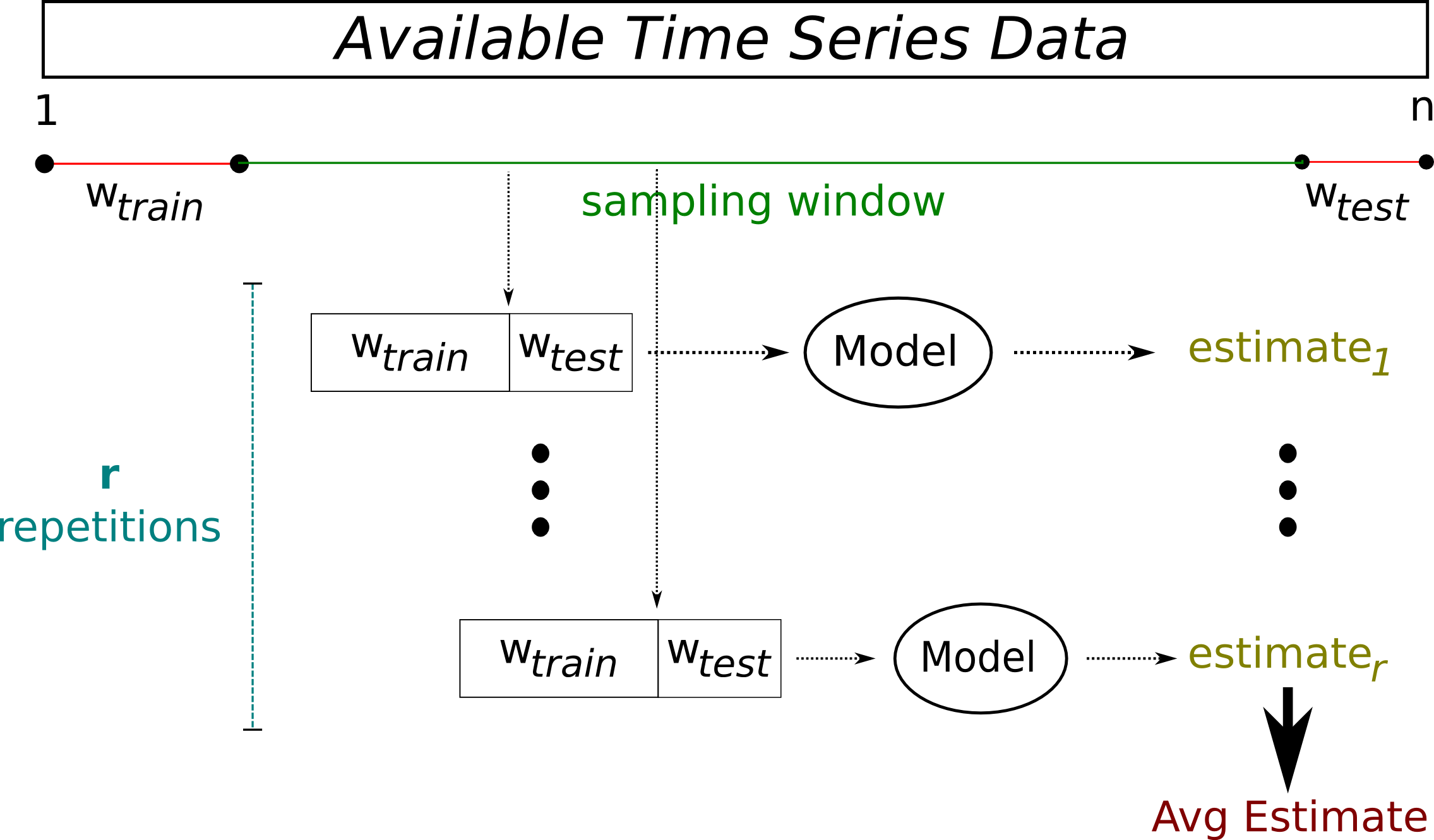}
  \caption{The Monte Carlo estimation methodology.}
  \label{fig:MC}
\end{figure}

Monte Carlo experiments are similar to random sub-sampling (or repeated
Holdout) in the sense that they consist of repeating a learning +
testing cycle several times using different and eventually overlapping data samples. The main
difference lies on the way the samples are obtained. In Monte Carlo
experiments the original order of the observations is respected and
train and test splits are obtained such that the testing samples
appear ``after'' the training samples, thus being the methodology of
choice when you are comparing time series forecasting models. The idea
of Monte Carlo experiments is the following: (i) given a data set
spanning from time $t_1$ till time $t_N$, (ii) given a training set
time interval size $w_{train}$ and a test set time interval size $w_{test}$, such
that $w_{train}+w_{test} < N$, (iii) Monte Carlo experiments generate $r$ random time
points from the interval $[t_{1+w_{train}},t_{N-w_{test}}]$, and then (iv) for each
of these $r$ time points they generate a training set with data in the
interval $[t_{r-w_{train}+1},t_{r}]$ and a test set with data in the interval
$[t_{r+1},t_{r+w_{test}}]$. Using this process $r$ train+test cycles are
carried out using the user-supplied workflow function, and the
experiment estimates result from the average of the $r$ scores as
usual. The overall process is depicted in Figure~\ref{fig:MC}.

To carry out this type of experiments in our infra-structure we can
include an S4 object of class \textbf{MonteCarlo} in the third
argument of function \texttt{performanceEstimation()}.

The function \texttt{MonteCarlo()} can be used as a constructor of
objects of class \textbf{MonteCarlo}. It accepts the following
arguments:

\begin{description}
\item[nReps] - the number of repetitions of the Monte Carlo experiment (default is $10$)
\item[szTrain] - the percentage (a number between 0 and 1) or the actual number of cases to use in the training samples (default is $0.25$)
\item[szTest] - the percentage (a number between 0 and 1) or the actual  number of cases to use in the test samples (default is $0.25$)
\item[seed] - the random number generator seed to use (default is $1234$)
\item[dataSplits] - a list containing user-supplied data splits
  for each of the repetitions (check the help page of the
  class for further details). This parameter defaults to
  \texttt{NULL}, i.e. no user-supplied splits, they are decided
  internally by the infra-structure.
\end{description}

The following is a small illustrative example using the quotes of the
SP500 index. This example compares two random forests with 500
regression trees, one applied in a standard way, and the other using
a sliding window with a relearn step of 5 days. The experiment
uses 10 repetitions of a train+test cycle using 50\% of the available
data for training and 25\% for testing.

\begin{knitrout}\small
\definecolor{shadecolor}{rgb}{0.969, 0.969, 0.969}\color{fgcolor}\begin{kframe}
\begin{alltt}
\hlkwd{library}\hlstd{(quantmod)}
\hlkwd{library}\hlstd{(randomForest)}
\hlkwd{getSymbols}\hlstd{(}\hlstr{'^GSPC'}\hlstd{,}\hlkwc{from}\hlstd{=}\hlstr{'2008-01-01'}\hlstd{,}\hlkwc{to}\hlstd{=}\hlstr{'2012-12-31'}\hlstd{)}
\hlstd{data.model} \hlkwb{<-} \hlkwd{specifyModel}\hlstd{(}
  \hlkwd{Next}\hlstd{(}\hlnum{100}\hlopt{*}\hlkwd{Delt}\hlstd{(}\hlkwd{Ad}\hlstd{(GSPC)))} \hlopt{~} \hlkwd{Delt}\hlstd{(}\hlkwd{Ad}\hlstd{(GSPC),}\hlkwc{k}\hlstd{=}\hlnum{1}\hlopt{:}\hlnum{10}\hlstd{))}
\hlstd{data} \hlkwb{<-} \hlkwd{modelData}\hlstd{(data.model)}
\hlkwd{colnames}\hlstd{(data)[}\hlnum{1}\hlstd{]} \hlkwb{<-} \hlstr{'PercVarClose'}
\hlstd{spExp} \hlkwb{<-} \hlkwd{performanceEstimation}\hlstd{(}
  \hlkwd{PredTask}\hlstd{(PercVarClose} \hlopt{~} \hlstd{.,data,}\hlstr{'SP500_2012'}\hlstd{),}
  \hlkwd{c}\hlstd{(}\hlkwd{Workflow}\hlstd{(}\hlstr{'standardWF'}\hlstd{,}\hlkwc{wfID}\hlstd{=}\hlstr{"standRF"}\hlstd{,}
             \hlkwc{learner}\hlstd{=}\hlstr{'randomForest'}\hlstd{,}\hlkwc{learner.pars}\hlstd{=}\hlkwd{list}\hlstd{(}\hlkwc{ntree}\hlstd{=}\hlnum{500}\hlstd{)),}
    \hlkwd{Workflow}\hlstd{(}\hlstr{'timeseriesWF'}\hlstd{,}\hlkwc{wfID}\hlstd{=}\hlstr{"slideRF"}\hlstd{,}
             \hlkwc{learner}\hlstd{=}\hlstr{'randomForest'}\hlstd{,}
             \hlkwc{learner.pars}\hlstd{=}\hlkwd{list}\hlstd{(}\hlkwc{ntree}\hlstd{=}\hlnum{500}\hlstd{,}\hlkwc{relearn.step}\hlstd{=}\hlnum{30}\hlstd{))}
   \hlstd{),}
  \hlkwd{EstimationTask}\hlstd{(}\hlkwc{metrics}\hlstd{=}\hlstr{"theil"}\hlstd{,}
                 \hlkwc{method}\hlstd{=}\hlkwd{MonteCarlo}\hlstd{(}\hlkwc{nReps}\hlstd{=}\hlnum{10}\hlstd{,}\hlkwc{szTrain}\hlstd{=}\hlnum{0.5}\hlstd{,}\hlkwc{szTest}\hlstd{=}\hlnum{0.25}\hlstd{)))}
\end{alltt}
\end{kframe}
\end{knitrout}

Note that in the above example we have not tried any variants of the two workflows that are applied to the task. This means that we have used directly the \texttt{Workflow} constructor to create our workflow. Note also the use of the \texttt{wfID} parameter of this constructor to allow you to give a particular workflow ID to some approach.

\section{Statistical Significance of Differences}

The estimation methodologies that we have presented in the previous
sections allow the user to obtain estimates of the mean predictive
performance of different workflows or variants of these workflows, on
different predictive tasks. We have seen that by applying the
\texttt{summary} method to the objects resulting from the estimation experiments
we can obtain the average performance for each candidate workflow on
each task. These numbers are estimates of the expected mean
performance of the workflows on the respective tasks. Being estimates,
the obvious next question is to check whether the observed differences
in performance between the workflows are statistically
significant. More formally, we want to know that confidence level of rejecting the null hypothesis that the difference between the estimated averages is zero.

That is the goal of the function
\texttt{pairedComparisons()}. This function implements a series of statistical hypothesis tests that can be used in different types of setups. We follow the general recommendations of Demsar~\cite{Dem06}. The function \texttt{pairedComparisons()} implements a series of statistical tests that can be used under different conditions. The function calculates them all (whenever possible) and it is up to the user to use the ones that are more adequate to answer her/his research questions (see Demsar~\cite{Dem06} for a series of guidelines on how to proceed).

Function \texttt{pairedComparisons()} returns a list with as many components as there are metrics estimated. For each metric another list includes the different tests that are carried out as well as several other information. Namely, for each error metric the returned list includes the following components:

\begin{itemize}
\item \textbf{setup} - with generic information on the estimation task
  \item \textbf{avgScores} - a matrix with the average scores on the metric of each workflow on each task
  \item \textbf{medScores} - another matrix with information like the previous one, but instead of the average we get the median scores
  \item \textbf{rks} - a matrix with the rank positions corresponding to the average scores
  \item \textbf{avgRksWfs} - a vector with the average of the above ranks across all tasks
  \item \textbf{t.test} - a list containing information concerning paired \textit{t} tests
  \item \textbf{WilcoxonSignedRank.test} - a list containing information concerning paired \textit{Wilcoxon Signed Rank} tests
  \item \textbf{F.test} - a list containing information concerning the \textit{F} test
  \item \textbf{Nemenyi.test} - a list containing information concerning the post-hoc \textit{Nemenyi} test
  \item \textbf{BonferroniDunn.test} - a list containing information concerning the post-hoc \textit{Bonferroni-Dunn} test
\end{itemize}

For both the \textit{t} and \textit{Wilcoxon Signed Rank} tests the components contain and array with 3 dimensions, the third dimension being the task, which means that for each task you get a matrix of results of paired comparisons. Namely, paired comparisons between each workflow and the baseline workflow. Each row of the matrix contains the average (or median in the case of Wilcoxon) score of the respective workflow, the difference between this score and the score of the baseline workflow, and the \textit{p} value associated with the hypothesis that this difference is distributed around mean (median in Wilcoxon). Please note that although our function calculates the results of the \textit{t} test, this is not recommended for most settings given the lack of independence of the values on each iteration (for instance on \textit{k}-fold cross validation each of the \textit{k} scores is obtained with training sets that have a string overlap).

The other 3 tests can be used in the following two settings: (i) test if there are significant differences among any pair of workflows; and (ii) test the significance of the differences between all workflows against a baseline. In both of these two settings we should start by checking the results of the  \textit{F} test to check if we can reject the null hypothesis that the average ranks of all workflows are equivalent. In case this hypothesis is rejected (component \texttt{rejNull} of the list \texttt{F.test}) we can move to the post-hoc tests. In the setup (i) we should use the \textit{Nemenyi} test to find the critical difference among ranks above which we can say that the respective difference is statistically significant. The \textit{Nemenyi.test} component is a list with components containing this information (critical difference value, ranking differences among all pairs of workflows, and whether these differences are or not significant). Under the setup (ii) the post-hoc test that should be used is the \textit{Bonferroni-Dunn} test. In this case we just want to check if the difference between the average rank of each workflow and that of the baseline is or not significant (thus much less paired comparisons). The result is again given as a list where we have information on the critical difference, the baseline workflow name, the vector of average differences and whether these differences are different or not.

Let us see a concrete example of using this statistical tests. Suppose we want to apply several variants of an SVM to three classification tasks and we want to check if there are significant differences among them. The following test carries out the performance estimation task:

\begin{knitrout}\scriptsize
\definecolor{shadecolor}{rgb}{0.969, 0.969, 0.969}\color{fgcolor}\begin{kframe}
\begin{alltt}
\hlkwd{library}\hlstd{(e1071)}
\hlkwd{data}\hlstd{(PimaIndiansDiabetes,}\hlkwc{package}\hlstd{=}\hlstr{'mlbench'}\hlstd{)}
\hlkwd{data}\hlstd{(iris)}
\hlkwd{data}\hlstd{(Glass,}\hlkwc{package}\hlstd{=}\hlstr{"mlbench"}\hlstd{)}
\hlstd{res} \hlkwb{<-} \hlkwd{performanceEstimation}\hlstd{(}
    \hlkwd{c}\hlstd{(}\hlkwd{PredTask}\hlstd{(diabetes} \hlopt{~} \hlstd{.,PimaIndiansDiabetes,}\hlstr{"Pima"}\hlstd{),}
      \hlkwd{PredTask}\hlstd{(Type} \hlopt{~} \hlstd{., Glass),}
      \hlkwd{PredTask}\hlstd{(Species} \hlopt{~} \hlstd{.,iris)),}
    \hlkwd{workflowVariants}\hlstd{(}\hlkwc{learner}\hlstd{=}\hlstr{"svm"}\hlstd{,}
                     \hlkwc{learner.pars}\hlstd{=}\hlkwd{list}\hlstd{(}\hlkwc{cost}\hlstd{=}\hlnum{1}\hlopt{:}\hlnum{5}\hlstd{,}\hlkwc{gamma}\hlstd{=}\hlkwd{c}\hlstd{(}\hlnum{0.1}\hlstd{,}\hlnum{0.01}\hlstd{,}\hlnum{0.001}\hlstd{))),}
    \hlkwd{EstimationTask}\hlstd{(}\hlkwc{metrics}\hlstd{=}\hlstr{"err"}\hlstd{,}\hlkwc{method}\hlstd{=}\hlkwd{CV}\hlstd{()))}
\end{alltt}
\end{kframe}
\end{knitrout}

The object \texttt{res} can be passed to \texttt{pairedComparisons()} to calculate the tests described before:

\begin{knitrout}\footnotesize
\definecolor{shadecolor}{rgb}{0.969, 0.969, 0.969}\color{fgcolor}\begin{kframe}
\begin{alltt}
\hlstd{pres} \hlkwb{<-} \hlkwd{pairedComparisons}\hlstd{(res)}
\end{alltt}
\end{kframe}
\end{knitrout}

As we have several workflows being compared on different tasks we will start by checking the results of the Friedman \textit{F} test:

\begin{knitrout}\footnotesize
\definecolor{shadecolor}{rgb}{0.969, 0.969, 0.969}\color{fgcolor}\begin{kframe}
\begin{alltt}
\hlstd{pres}\hlopt{$}\hlstd{err}\hlopt{$}\hlstd{F.test}
\end{alltt}
\begin{verbatim}
## $chi
## [1] 18.575
## 
## $FF
## [1] 1.585912
## 
## $critVal
## [1] 0.8892237
## 
## $rejNull
## [1] TRUE
\end{verbatim}
\end{kframe}
\end{knitrout}

We can see that the null hypothesis that the average ranks of the workflows are equivalent can be rejected and thus we can proceed to the post-hoc test. In case we wish the check the hypothesis that there are statistically significant differences among the workflows we should proceed with a Nemenyi post-hoc test. The results of this test can be inspected as follows:

\begin{knitrout}\tiny
\definecolor{shadecolor}{rgb}{0.969, 0.969, 0.969}\color{fgcolor}\begin{kframe}
\begin{alltt}
\hlstd{pres}\hlopt{$}\hlstd{err}\hlopt{$}\hlstd{Nemenyi.test}
\end{alltt}
\begin{verbatim}
## $critDif
## [1] 12.38302
## 
## $rkDifs
##            svm.v1    svm.v2    svm.v3    svm.v4    svm.v5    svm.v6    svm.v7    svm.v8
## svm.v1  0.0000000 1.5000000 1.0000000 1.8333333 1.5000000 0.6666667 0.8333333 0.1666667
## svm.v2  1.5000000 0.0000000 0.5000000 0.3333333 0.0000000 2.1666667 2.3333333 1.3333333
## svm.v3  1.0000000 0.5000000 0.0000000 0.8333333 0.5000000 1.6666667 1.8333333 0.8333333
## svm.v4  1.8333333 0.3333333 0.8333333 0.0000000 0.3333333 2.5000000 2.6666667 1.6666667
## svm.v5  1.5000000 0.0000000 0.5000000 0.3333333 0.0000000 2.1666667 2.3333333 1.3333333
## svm.v6  0.6666667 2.1666667 1.6666667 2.5000000 2.1666667 0.0000000 0.1666667 0.8333333
## svm.v7  0.8333333 2.3333333 1.8333333 2.6666667 2.3333333 0.1666667 0.0000000 1.0000000
## svm.v8  0.1666667 1.3333333 0.8333333 1.6666667 1.3333333 0.8333333 1.0000000 0.0000000
## svm.v9  1.3333333 0.1666667 0.3333333 0.5000000 0.1666667 2.0000000 2.1666667 1.1666667
## svm.v10 2.8333333 1.3333333 1.8333333 1.0000000 1.3333333 3.5000000 3.6666667 2.6666667
## svm.v11 7.6666667 9.1666667 8.6666667 9.5000000 9.1666667 7.0000000 6.8333333 7.8333333
## svm.v12 6.6666667 8.1666667 7.6666667 8.5000000 8.1666667 6.0000000 5.8333333 6.8333333
## svm.v13 2.3333333 3.8333333 3.3333333 4.1666667 3.8333333 1.6666667 1.5000000 2.5000000
## svm.v14 1.0000000 2.5000000 2.0000000 2.8333333 2.5000000 0.3333333 0.1666667 1.1666667
## svm.v15 1.0000000 2.5000000 2.0000000 2.8333333 2.5000000 0.3333333 0.1666667 1.1666667
##            svm.v9   svm.v10   svm.v11  svm.v12  svm.v13   svm.v14   svm.v15
## svm.v1  1.3333333  2.833333  7.666667 6.666667 2.333333 1.0000000 1.0000000
## svm.v2  0.1666667  1.333333  9.166667 8.166667 3.833333 2.5000000 2.5000000
## svm.v3  0.3333333  1.833333  8.666667 7.666667 3.333333 2.0000000 2.0000000
## svm.v4  0.5000000  1.000000  9.500000 8.500000 4.166667 2.8333333 2.8333333
## svm.v5  0.1666667  1.333333  9.166667 8.166667 3.833333 2.5000000 2.5000000
## svm.v6  2.0000000  3.500000  7.000000 6.000000 1.666667 0.3333333 0.3333333
## svm.v7  2.1666667  3.666667  6.833333 5.833333 1.500000 0.1666667 0.1666667
## svm.v8  1.1666667  2.666667  7.833333 6.833333 2.500000 1.1666667 1.1666667
## svm.v9  0.0000000  1.500000  9.000000 8.000000 3.666667 2.3333333 2.3333333
## svm.v10 1.5000000  0.000000 10.500000 9.500000 5.166667 3.8333333 3.8333333
## svm.v11 9.0000000 10.500000  0.000000 1.000000 5.333333 6.6666667 6.6666667
## svm.v12 8.0000000  9.500000  1.000000 0.000000 4.333333 5.6666667 5.6666667
## svm.v13 3.6666667  5.166667  5.333333 4.333333 0.000000 1.3333333 1.3333333
## svm.v14 2.3333333  3.833333  6.666667 5.666667 1.333333 0.0000000 0.0000000
## svm.v15 2.3333333  3.833333  6.666667 5.666667 1.333333 0.0000000 0.0000000
## 
## $signifDifs
##         svm.v1 svm.v2 svm.v3 svm.v4 svm.v5 svm.v6 svm.v7 svm.v8 svm.v9 svm.v10 svm.v11
## svm.v1   FALSE  FALSE  FALSE  FALSE  FALSE  FALSE  FALSE  FALSE  FALSE   FALSE   FALSE
## svm.v2   FALSE  FALSE  FALSE  FALSE  FALSE  FALSE  FALSE  FALSE  FALSE   FALSE   FALSE
## svm.v3   FALSE  FALSE  FALSE  FALSE  FALSE  FALSE  FALSE  FALSE  FALSE   FALSE   FALSE
## svm.v4   FALSE  FALSE  FALSE  FALSE  FALSE  FALSE  FALSE  FALSE  FALSE   FALSE   FALSE
## svm.v5   FALSE  FALSE  FALSE  FALSE  FALSE  FALSE  FALSE  FALSE  FALSE   FALSE   FALSE
## svm.v6   FALSE  FALSE  FALSE  FALSE  FALSE  FALSE  FALSE  FALSE  FALSE   FALSE   FALSE
## svm.v7   FALSE  FALSE  FALSE  FALSE  FALSE  FALSE  FALSE  FALSE  FALSE   FALSE   FALSE
## svm.v8   FALSE  FALSE  FALSE  FALSE  FALSE  FALSE  FALSE  FALSE  FALSE   FALSE   FALSE
## svm.v9   FALSE  FALSE  FALSE  FALSE  FALSE  FALSE  FALSE  FALSE  FALSE   FALSE   FALSE
## svm.v10  FALSE  FALSE  FALSE  FALSE  FALSE  FALSE  FALSE  FALSE  FALSE   FALSE   FALSE
## svm.v11  FALSE  FALSE  FALSE  FALSE  FALSE  FALSE  FALSE  FALSE  FALSE   FALSE   FALSE
## svm.v12  FALSE  FALSE  FALSE  FALSE  FALSE  FALSE  FALSE  FALSE  FALSE   FALSE   FALSE
## svm.v13  FALSE  FALSE  FALSE  FALSE  FALSE  FALSE  FALSE  FALSE  FALSE   FALSE   FALSE
## svm.v14  FALSE  FALSE  FALSE  FALSE  FALSE  FALSE  FALSE  FALSE  FALSE   FALSE   FALSE
## svm.v15  FALSE  FALSE  FALSE  FALSE  FALSE  FALSE  FALSE  FALSE  FALSE   FALSE   FALSE
##         svm.v12 svm.v13 svm.v14 svm.v15
## svm.v1    FALSE   FALSE   FALSE   FALSE
## svm.v2    FALSE   FALSE   FALSE   FALSE
## svm.v3    FALSE   FALSE   FALSE   FALSE
## svm.v4    FALSE   FALSE   FALSE   FALSE
## svm.v5    FALSE   FALSE   FALSE   FALSE
## svm.v6    FALSE   FALSE   FALSE   FALSE
## svm.v7    FALSE   FALSE   FALSE   FALSE
## svm.v8    FALSE   FALSE   FALSE   FALSE
## svm.v9    FALSE   FALSE   FALSE   FALSE
## svm.v10   FALSE   FALSE   FALSE   FALSE
## svm.v11   FALSE   FALSE   FALSE   FALSE
## svm.v12   FALSE   FALSE   FALSE   FALSE
## svm.v13   FALSE   FALSE   FALSE   FALSE
## svm.v14   FALSE   FALSE   FALSE   FALSE
## svm.v15   FALSE   FALSE   FALSE   FALSE
\end{verbatim}
\end{kframe}
\end{knitrout}

This is obviously a too extensive set of information though the conclusions are centered in the \texttt{signifDifs} component. An even better alternative can be obtained through CD diagrams, which we have implemented in function \texttt{CDdiagram.Nemenyi} whose result is shown in Figure~\ref{fig:cdn}.

\begin{knitrout}\scriptsize
\definecolor{shadecolor}{rgb}{0.969, 0.969, 0.969}\color{fgcolor}\begin{kframe}
\begin{alltt}
\hlkwd{CDdiagram.Nemenyi}\hlstd{(pres)}
\end{alltt}
\end{kframe}\begin{figure}[]

{\centering \includegraphics[width=0.9\textwidth]{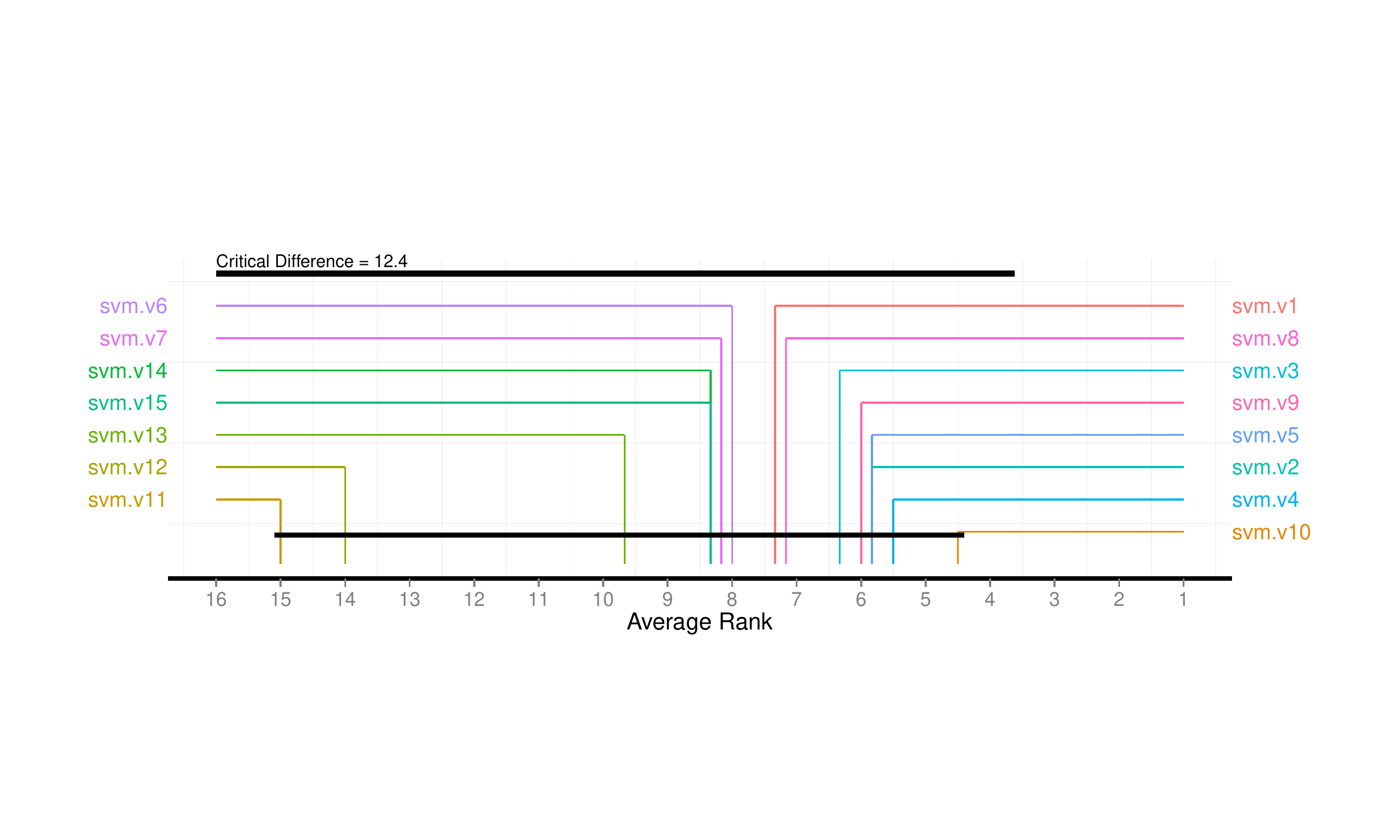} 

}

\caption[The CD diagram of the Nemenyi post-hoc test]{The CD diagram of the Nemenyi post-hoc test.\label{fig:cdn}}
\end{figure}

\end{knitrout}

At the top left of the diagram you can see a horizontal line with the critical difference for the differences between average ranks to be considered statistically significant. Then each workflow is represented by a line with origin in its respective average rank (in the X axis). If the lines of any pair of workflows are connected by a horizontal black line, it means that the difference between them is not statistically significant. In the diagram of Figure~\ref{fig:cdn} we can observe that all lines are connected meaning that none of the paired differences between all workflows can be seen as statistically significant.

\section{Larger Examples}

The main advantage of the infra-structure we are proposing is to
automate large scale performance estimation experiments. It is on these very
large setups that the use of the infra-structure saves more time to
the user. However, in these contexts the objects resulting from the
estimation process are very large and some of the tools we have shown before
for exploring the results may produce over-cluttered output. In
effect, if you have an experiment involving dozens of predictive tasks
and eventually hundreds of workflow variants being compared on several
evaluation metrics, doing a plot of the resulting object is simply not
possible as the graph will be unreadable. This section illustrates
some of these cases and presents some solutions to overcome the
difficulties they bring.

Extremely large experiments may take days or weeks to complete,
depending on the available hardware. In this context, it may not be
wise to run the experiments on a single call to the
\texttt{performanceEstimation()} function because if something goes
wrong in the middle you may loose lots of work/time. Using the random
number generation seeds that are available in all experimental
settings objects we can split the experiments in several calls and
still ensure that the same data folds are used in all
estimation experiments. If the seeding process is not enough due to the usage of different hardware for instance, then you may still resort to the user supplied data splits to make sure all methods are compared on the same data. We will see that when all experiments are
finished we will be able to merge the objects of each call into a
single object as if we had issued a single call. 

Another way of improving performance on large experiments is through parallel computation. \PE includes facilities to run experiment in parallel through the existing parallel back-ends in R. The simplest of options, that has the advantage of not requiring you to know anything at all about parallel computation in R, is to call the \texttt{performanceEstimation} function with a fourth parameter that will make all experiments to run in parallel on the different cores of any standard multicore computer. The following is a simple example of achieving this effect:

\begin{knitrout}\small
\definecolor{shadecolor}{rgb}{0.969, 0.969, 0.969}\color{fgcolor}\begin{kframe}
\begin{alltt}
\hlkwd{library}\hlstd{(performanceEstimation)}
\hlkwd{library}\hlstd{(e1071)}
\hlkwd{data}\hlstd{(Satellite,}\hlkwc{package}\hlstd{=}\hlstr{"mlbench"}\hlstd{)}
\hlstd{pres} \hlkwb{<-} \hlkwd{performanceEstimation}\hlstd{(}
    \hlkwd{PredTask}\hlstd{(classes} \hlopt{~} \hlstd{.,Satellite),}
    \hlkwd{Workflow}\hlstd{(}\hlkwc{learner}\hlstd{=}\hlstr{"svm"}\hlstd{),}
    \hlkwd{EstimationTask}\hlstd{(}\hlstr{"err"}\hlstd{,}\hlkwd{CV}\hlstd{()),}
    \hlkwc{cluster}\hlstd{=}\hlnum{TRUE}
    \hlstd{)}
\end{alltt}
\begin{verbatim}
## 
## 
## ##### PERFORMANCE ESTIMATION USING  CROSS VALIDATION  #####
## 
## ** PREDICTIVE TASK :: Satellite.classes
## 
## ++ MODEL/WORKFLOW :: svm 
## cvEstimates: Running in parallel with 2 worker(s)
## Task for estimating  err  using
##  1 x 10 - Fold Cross Validation
## 	 Run with seed =  1234
\end{verbatim}
\end{kframe}
\end{knitrout}

By adding this extra parameter the lower level functions that implement the estimation procedures (in this case cross validation but the concept is applicable to any of the other methods), will automatically create a (local) parallel back-end using half of the cores of your computer. On reasonably large data sets this simple extra setting will typically lead to cut execution times by a significant amount. 

Other more complex parallel settings are also possible but they require you to know how to create the clusters before calling our function. For instance, you could create a cluster (using function \texttt{makeCluster} of package \textbf{parallel}) that involved several computers. The object of class \texttt{cluster} that results from these steps should then be sent through parameter \texttt{cluster} of our \texttt{performanceEstimation} function so that the function can take advantage of this parallel back-end. The following is a simple illustrative example:

\begin{knitrout}\small
\definecolor{shadecolor}{rgb}{0.969, 0.969, 0.969}\color{fgcolor}\begin{kframe}
\begin{alltt}
\hlkwd{library}\hlstd{(parallel)}
\hlstd{myclust} \hlkwb{<-} \hlkwd{makeCluster}\hlstd{(}\hlkwd{c}\hlstd{(}\hlstr{"localhost"}\hlstd{,}\hlstr{"192.168.2.10"}\hlstd{,}\hlstr{"192.168.2.13"}\hlstd{),}\hlstr{"SOCK"}\hlstd{)}
\hlkwd{library}\hlstd{(performanceEstimation)}
\hlkwd{library}\hlstd{(e1071)}
\hlkwd{data}\hlstd{(Satellite,}\hlkwc{package}\hlstd{=}\hlstr{"mlbench"}\hlstd{)}
\hlstd{pres} \hlkwb{<-} \hlkwd{performanceEstimation}\hlstd{(}
    \hlkwd{PredTask}\hlstd{(classes} \hlopt{~} \hlstd{.,Satellite),}
    \hlkwd{Workflow}\hlstd{(}\hlkwc{learner}\hlstd{=}\hlstr{"svm"}\hlstd{),}
    \hlkwd{EstimationTask}\hlstd{(}\hlstr{"err"}\hlstd{,}\hlkwd{CV}\hlstd{()),}
    \hlkwc{cluster}\hlstd{=myclust}
    \hlstd{)}
\hlkwd{stopCluster}\hlstd{(myclust)}
\end{alltt}
\end{kframe}
\end{knitrout}

Please note that there are a few assumptions on the above example, namely concerning access of your username in localhost to the remote machines whose IP's are given. This means that the username exists on all machines and that appropriate SSH keys have been set so that no passwords are required for accessing the remote machines.

\vspace*{0.5cm}

Let us now focus on the issue of splitting a large experiment in partial calls to \texttt{performanceEstimation()} and at the end merge the separate results into a single object as if the experiments were run all together. Please note that further efficiency gains could be achieved if some parallel setup involving different computers as the one described above was used. The following example addresses several regression tasks using several workflows that are variants of different regression algorithms. We illustrate how to create the different \textbf{PredTask} objects programmatically,  how to create the different workflow variants, and how to call \texttt{performanceEstimation()} with different task/workflow combinations storing the intermediate objects on temporary files for later reading and merging,

\begin{knitrout}\small
\definecolor{shadecolor}{rgb}{0.969, 0.969, 0.969}\color{fgcolor}\begin{kframe}
\begin{alltt}
\hlkwd{library}\hlstd{(performanceEstimation)}
\hlkwd{library}\hlstd{(e1071)}
\hlkwd{library}\hlstd{(randomForest)}

\hlkwd{data}\hlstd{(algae,}\hlkwc{package}\hlstd{=}\hlstr{"DMwR"}\hlstd{)}
\hlstd{DSs} \hlkwb{<-} \hlkwd{sapply}\hlstd{(}\hlkwd{names}\hlstd{(algae)[}\hlnum{12}\hlopt{:}\hlnum{18}\hlstd{],}
         \hlkwa{function}\hlstd{(}\hlkwc{x}\hlstd{,}\hlkwc{names.attrs}\hlstd{) \{}
           \hlstd{f} \hlkwb{<-} \hlkwd{as.formula}\hlstd{(}\hlkwd{paste}\hlstd{(x,}\hlstr{"~ ."}\hlstd{))}
           \hlkwd{PredTask}\hlstd{(f,algae[,}\hlkwd{c}\hlstd{(names.attrs,x)],x,}\hlkwc{copy}\hlstd{=}\hlnum{TRUE}\hlstd{)}
         \hlstd{\},}
         \hlkwd{names}\hlstd{(algae)[}\hlnum{1}\hlopt{:}\hlnum{11}\hlstd{])}

\hlstd{WFs} \hlkwb{<-} \hlkwd{list}\hlstd{()}
\hlstd{WFs}\hlopt{$}\hlstd{svm} \hlkwb{<-} \hlkwd{list}\hlstd{(}\hlkwc{learner.pars}\hlstd{=}\hlkwd{list}\hlstd{(}\hlkwc{cost}\hlstd{=}\hlkwd{c}\hlstd{(}\hlnum{10}\hlstd{,}\hlnum{150}\hlstd{,}\hlnum{300}\hlstd{),}
                                  \hlkwc{gamma}\hlstd{=}\hlkwd{c}\hlstd{(}\hlnum{0.01}\hlstd{,}\hlnum{0.001}\hlstd{),}
                                  \hlkwc{epsilon}\hlstd{=}\hlkwd{c}\hlstd{(}\hlnum{0.1}\hlstd{,}\hlnum{0.05}\hlstd{)),}
                \hlkwc{pre}\hlstd{=}\hlstr{"centralImp"}\hlstd{,}\hlkwc{post}\hlstd{=}\hlstr{"na2central"}\hlstd{)}
\hlstd{WFs}\hlopt{$}\hlstd{randomForest} \hlkwb{<-} \hlkwd{list}\hlstd{(}\hlkwc{learner.pars}\hlstd{=}\hlkwd{list}\hlstd{(}\hlkwc{mtry}\hlstd{=}\hlkwd{c}\hlstd{(}\hlnum{5}\hlstd{,}\hlnum{7}\hlstd{),}
                                           \hlkwc{ntree}\hlstd{=}\hlkwd{c}\hlstd{(}\hlnum{500}\hlstd{,}\hlnum{750}\hlstd{,}\hlnum{1500}\hlstd{)),}
                         \hlkwc{pre}\hlstd{=}\hlstr{"centralImp"}\hlstd{)}

\hlkwa{for}\hlstd{(d} \hlkwa{in} \hlkwd{seq_along}\hlstd{(DSs)) \{}
  \hlkwa{for}\hlstd{(w} \hlkwa{in} \hlkwd{names}\hlstd{(WFs)) \{}
    \hlstd{resObj} \hlkwb{<-} \hlkwd{paste}\hlstd{(}\hlkwd{names}\hlstd{(DSs)[d],w,}\hlstr{'Res'}\hlstd{,}\hlkwc{sep}\hlstd{=}\hlstr{''}\hlstd{)}
    \hlkwd{assign}\hlstd{(resObj,}
           \hlkwd{performanceEstimation}\hlstd{(}
                  \hlstd{DSs[d],}
                  \hlkwd{c}\hlstd{(}
                    \hlkwd{do.call}\hlstd{(}\hlstr{'workflowVariants'}\hlstd{,}
                            \hlkwd{c}\hlstd{(}\hlkwd{list}\hlstd{(}\hlkwc{learner}\hlstd{=w),WFs[[w]]))}
                    \hlstd{),}
                   \hlkwd{EstimationTask}\hlstd{(}\hlkwc{metrics}\hlstd{=}\hlkwd{c}\hlstd{(}\hlstr{'mse'}\hlstd{,}\hlstr{'mae'}\hlstd{),}\hlkwc{method}\hlstd{=}\hlkwd{CV}\hlstd{(}\hlkwc{nReps}\hlstd{=}\hlnum{3}\hlstd{)),}
                   \hlkwc{cluster}\hlstd{=}\hlnum{TRUE}\hlstd{)}
           \hlstd{)}

    \hlkwd{save}\hlstd{(}\hlkwc{list}\hlstd{=resObj,}\hlkwc{file}\hlstd{=}\hlkwd{paste}\hlstd{(}\hlkwd{names}\hlstd{(DSs)[d],w,}\hlstr{'Rdata'}\hlstd{,}\hlkwc{sep}\hlstd{=}\hlstr{'.'}\hlstd{))}
  \hlstd{\}}
\hlstd{\}}
\end{alltt}
\end{kframe}
\end{knitrout}

The above code compares 12 SVM variants with 6 random forest variants,
on 7 algae blooms regression tasks, using $3\times 10-$fold cross
validation. Although this is not a very large experimental comparison
it still includes applying 18 different workflow variants on 7
different prediction tasks, 30 times, i.e. 3780 train+test
cycles. Instead of running all these experiments on a single call to
the function \texttt{performanceEstimation()} (which would obviously
still be possible), we have made different calls for each workflow
type (SVM and random forest) and for each predictive task. This
means that each call will run all variants of a certain workflow on a
certain predictive task. The result of each of these calls will be
assigned to an object with a name composed of the task name and
workflow learner (the \texttt{resObj} variable). In the end each of
these objects is saved on a file with a similar name, for future
loading and results analysis. For instance, in the end there will be a
file with name ``a1.svm.Rdata'' which contains an object of class
\textbf{ComparisonResults} named \texttt{a1svmRes}. This object
contains the MSE and MAE estimated scores of the SVM variants on the
task of predicting the target variable ``a1'' (one of the seven algae
in this data set).

Later on, after the above experiment has finished you can load the saved objects back 
into R and moreover, join them into a single object, as shown below:

\begin{knitrout}\footnotesize
\definecolor{shadecolor}{rgb}{0.969, 0.969, 0.969}\color{fgcolor}\begin{kframe}
\begin{alltt}
\hlstd{nD} \hlkwb{<-} \hlkwd{paste}\hlstd{(}\hlstr{'a'}\hlstd{,}\hlnum{1}\hlopt{:}\hlnum{7}\hlstd{,}\hlkwc{sep}\hlstd{=}\hlstr{''}\hlstd{)}
\hlstd{nL} \hlkwb{<-} \hlkwd{c}\hlstd{(}\hlstr{'svm'}\hlstd{,}\hlstr{'randomForest'}\hlstd{)}
\hlstd{res} \hlkwb{<-} \hlkwa{NULL}
\hlkwa{for}\hlstd{(d} \hlkwa{in} \hlstd{nD) \{}
  \hlstd{resD} \hlkwb{<-} \hlkwa{NULL}
  \hlkwa{for}\hlstd{(l} \hlkwa{in} \hlstd{nL) \{}
    \hlkwd{load}\hlstd{(}\hlkwd{paste}\hlstd{(d,l,}\hlstr{'Rdata'}\hlstd{,}\hlkwc{sep}\hlstd{=}\hlstr{'.'}\hlstd{))}
    \hlstd{x} \hlkwb{<-} \hlkwd{get}\hlstd{(}\hlkwd{paste}\hlstd{(d,l,}\hlstr{'Res'}\hlstd{,}\hlkwc{sep}\hlstd{=}\hlstr{''}\hlstd{))}
    \hlstd{resD} \hlkwb{<-} \hlkwa{if} \hlstd{(}\hlkwd{is.null}\hlstd{(resD)) x} \hlkwa{else} \hlkwd{mergeEstimationRes}\hlstd{(resD,x,}\hlkwc{by}\hlstd{=}\hlstr{'workflows'}\hlstd{)}
  \hlstd{\}}
  \hlstd{res} \hlkwb{<-} \hlkwa{if} \hlstd{(}\hlkwd{is.null}\hlstd{(res)) resD} \hlkwa{else} \hlkwd{mergeEstimationRes}\hlstd{(res,resD,}\hlkwc{by}\hlstd{=}\hlstr{'tasks'}\hlstd{)}
\hlstd{\}}
\hlkwd{save}\hlstd{(res,}\hlkwc{file}\hlstd{=}\hlstr{'allResultsAlgae.Rdata'}\hlstd{)}
\end{alltt}
\end{kframe}
\end{knitrout}

The \texttt{mergeEstimationRes()}  function when applied to objects of class
\textbf{ComparisonResults} allows merging of these objects across different
dimensions. Namely, such objects have the individual scores of all
experiments spread across 3 dimensions:  the
metrics, the workflows and the tasks. The argument
\texttt{by} of the \texttt{mergeEstimationRes()} function allows you to specify how
to merge the given objects. The most common situations are: (i)
merging the results of different workflows over the same data sets -
you should use ``\texttt{by='workflows'}'', or (ii) merging the results
of the same workflows across different tasks - you should use
``\texttt{by='tasks'}''.

The following code can be used to check that the merging was OK, and
also to illustrate a few other utility functions whose purpose should
be obvious:

\begin{knitrout}\small
\definecolor{shadecolor}{rgb}{0.969, 0.969, 0.969}\color{fgcolor}\begin{kframe}
\begin{alltt}
\hlstd{res}
\end{alltt}
\begin{verbatim}
## 
## ==  Cross Validation Performance Estimation Experiment ==
## 
## Task for estimating  mse,mae  using
##  3 x 10 - Fold Cross Validation
## 	 Run with seed =  1234 
## 
##  18  workflows applied to  7  predictive tasks
\end{verbatim}
\begin{alltt}
\hlkwd{taskNames}\hlstd{(res)}
\end{alltt}
\begin{verbatim}
## [1] "a1" "a2" "a3" "a4" "a5" "a6" "a7"
\end{verbatim}
\begin{alltt}
\hlkwd{workflowNames}\hlstd{(res)}
\end{alltt}
\begin{verbatim}
##  [1] "svm.v1"          "svm.v2"          "svm.v3"          "svm.v4"         
##  [5] "svm.v5"          "svm.v6"          "svm.v7"          "svm.v8"         
##  [9] "svm.v9"          "svm.v10"         "svm.v11"         "svm.v12"        
## [13] "randomForest.v1" "randomForest.v2" "randomForest.v3" "randomForest.v4"
## [17] "randomForest.v5" "randomForest.v6"
\end{verbatim}
\begin{alltt}
\hlkwd{metricNames}\hlstd{(res)}
\end{alltt}
\begin{verbatim}
## [1] "mse" "mae"
\end{verbatim}
\end{kframe}
\end{knitrout}

With such large objects the most we can do is obtaining the best
scores or rankings of the workflows:

\begin{knitrout}\scriptsize
\definecolor{shadecolor}{rgb}{0.969, 0.969, 0.969}\color{fgcolor}\begin{kframe}
\begin{alltt}
\hlkwd{topPerformers}\hlstd{(res)}
\end{alltt}
\begin{verbatim}
## $a1
##            Workflow Estimate
## mse randomForest.v3  265.374
## mae randomForest.v1   11.066
## 
## $a2
##            Workflow Estimate
## mse randomForest.v5   94.463
## mae         svm.v11    5.917
## 
## $a3
##     Workflow Estimate
## mse   svm.v3   44.442
## mae  svm.v10    3.911
## 
## $a4
##            Workflow Estimate
## mse randomForest.v5   18.238
## mae         svm.v10    1.882
## 
## $a5
##            Workflow Estimate
## mse randomForest.v1   44.415
## mae          svm.v7    4.037
## 
## $a6
##     Workflow Estimate
## mse   svm.v6  111.757
## mae  svm.v12    5.548
## 
## $a7
##            Workflow Estimate
## mse randomForest.v3   26.208
## mae         svm.v10    2.383
\end{verbatim}
\end{kframe}
\end{knitrout}

\begin{knitrout}\scriptsize
\definecolor{shadecolor}{rgb}{0.969, 0.969, 0.969}\color{fgcolor}\begin{kframe}
\begin{alltt}
\hlkwd{rankWorkflows}\hlstd{(res)}
\end{alltt}
\begin{verbatim}
## $a1
## $a1$mse
##          Workflow Estimate
## 1 randomForest.v3 265.3743
## 2 randomForest.v5 265.6385
## 3 randomForest.v1 266.0238
## 4 randomForest.v6 272.3064
## 5 randomForest.v4 272.5888
## 
## $a1$mae
##          Workflow Estimate
## 1 randomForest.v1 11.06603
## 2 randomForest.v3 11.07444
## 3 randomForest.v5 11.08084
## 4 randomForest.v4 11.18655
## 5 randomForest.v2 11.19782
## 
## 
## $a2
## $a2$mse
##          Workflow Estimate
## 1 randomForest.v5 94.46308
## 2 randomForest.v6 94.88411
## 3 randomForest.v3 94.90805
## 4 randomForest.v1 94.95223
## 5 randomForest.v4 95.03046
## 
## $a2$mae
##   Workflow Estimate
## 1  svm.v11 5.917035
## 2   svm.v1 5.937923
## 3  svm.v10 5.940801
## 4   svm.v4 5.954024
## 5   svm.v5 5.967910
## 
## 
## $a3
## $a3$mse
##          Workflow Estimate
## 1          svm.v3 44.44234
## 2          svm.v2 44.54107
## 3          svm.v8 45.27262
## 4          svm.v9 45.31149
## 5 randomForest.v5 46.30568
## 
## $a3$mae
##   Workflow Estimate
## 1  svm.v10 3.911241
## 2   svm.v4 3.914820
## 3   svm.v7 3.943484
## 4   svm.v1 3.956006
## 5   svm.v5 4.006819
## 
## 
## $a4
## $a4$mse
##          Workflow Estimate
## 1 randomForest.v5 18.23780
## 2 randomForest.v3 18.25305
## 3 randomForest.v1 18.26055
## 4          svm.v4 18.36773
## 5 randomForest.v2 18.56495
## 
## $a4$mae
##   Workflow Estimate
## 1  svm.v10 1.881994
## 2   svm.v4 1.883301
## 3   svm.v7 1.983666
## 4   svm.v1 1.994250
## 5  svm.v11 2.022148
## 
## 
## $a5
## $a5$mse
##          Workflow Estimate
## 1 randomForest.v1 44.41519
## 2 randomForest.v3 44.43139
## 3 randomForest.v5 44.50338
## 4 randomForest.v4 45.50986
## 5 randomForest.v2 45.53312
## 
## $a5$mae
##   Workflow Estimate
## 1   svm.v7 4.036624
## 2  svm.v10 4.055948
## 3   svm.v4 4.061444
## 4   svm.v1 4.074267
## 5   svm.v5 4.077211
## 
## 
## $a6
## $a6$mse
##          Workflow Estimate
## 1          svm.v6 111.7569
## 2         svm.v12 113.1213
## 3          svm.v5 121.9274
## 4         svm.v11 123.1705
## 5 randomForest.v3 123.3406
## 
## $a6$mae
##   Workflow Estimate
## 1  svm.v12 5.548443
## 2   svm.v7 5.572169
## 3   svm.v6 5.587264
## 4   svm.v4 5.633257
## 5   svm.v1 5.643838
## 
## 
## $a7
## $a7$mse
##          Workflow Estimate
## 1 randomForest.v3 26.20759
## 2 randomForest.v5 26.21914
## 3 randomForest.v1 26.32753
## 4 randomForest.v2 26.44505
## 5 randomForest.v4 26.49540
## 
## $a7$mae
##   Workflow Estimate
## 1  svm.v10 2.382957
## 2   svm.v4 2.407783
## 3   svm.v7 2.419313
## 4   svm.v1 2.439576
## 5   svm.v6 2.439598
\end{verbatim}
\end{kframe}
\end{knitrout}

Notice that both \texttt{topPerformers()} and \texttt{rankWorkflows()}
assume that the evaluation metrics are to be minimized, i.e. they
assume the lower the better the scores. Still, both functions have a
parameter named \texttt{maxs} that accepts a vector with as many
Boolean values as there are evaluation metrics being estimated, which
you may use to indicate that some particular metric is to be maximized
and not minimized (the default). So for instance, if you had an
experiment where the 1st and 3rd metrics are to be minimized, whilst
the second is to be maximized, you could call these functions as
\texttt{rankWorkflows(resObj,maxs=c(F,T,F))}.

In order to obtain further results from these large objects one
usually proceeds by analyzing parts of the object, for instance
focusing on a particular task or metric, or even a subset of the
workflows. To facilitate this we can use the generic function
\texttt{subset()} that can also be applied to objects of class
\textbf{ComparisonResults}. An example of its use is given below, which results
in a graph of the performance of the different workflows in the
predictive task ``a1'', in terms of ``MAE'', which is show in
Figure~\ref{fig:maeA1}.

\begin{knitrout}\footnotesize
\definecolor{shadecolor}{rgb}{0.969, 0.969, 0.969}\color{fgcolor}\begin{kframe}
\begin{alltt}
\hlkwd{plot}\hlstd{(}\hlkwd{subset}\hlstd{(res,} \hlkwc{tasks}\hlstd{=}\hlstr{'a1'}\hlstd{,} \hlkwc{metrics}\hlstd{=}\hlstr{'mae'}\hlstd{))}
\end{alltt}
\end{kframe}\begin{figure}[]

{\centering \includegraphics[width=0.75\textwidth]{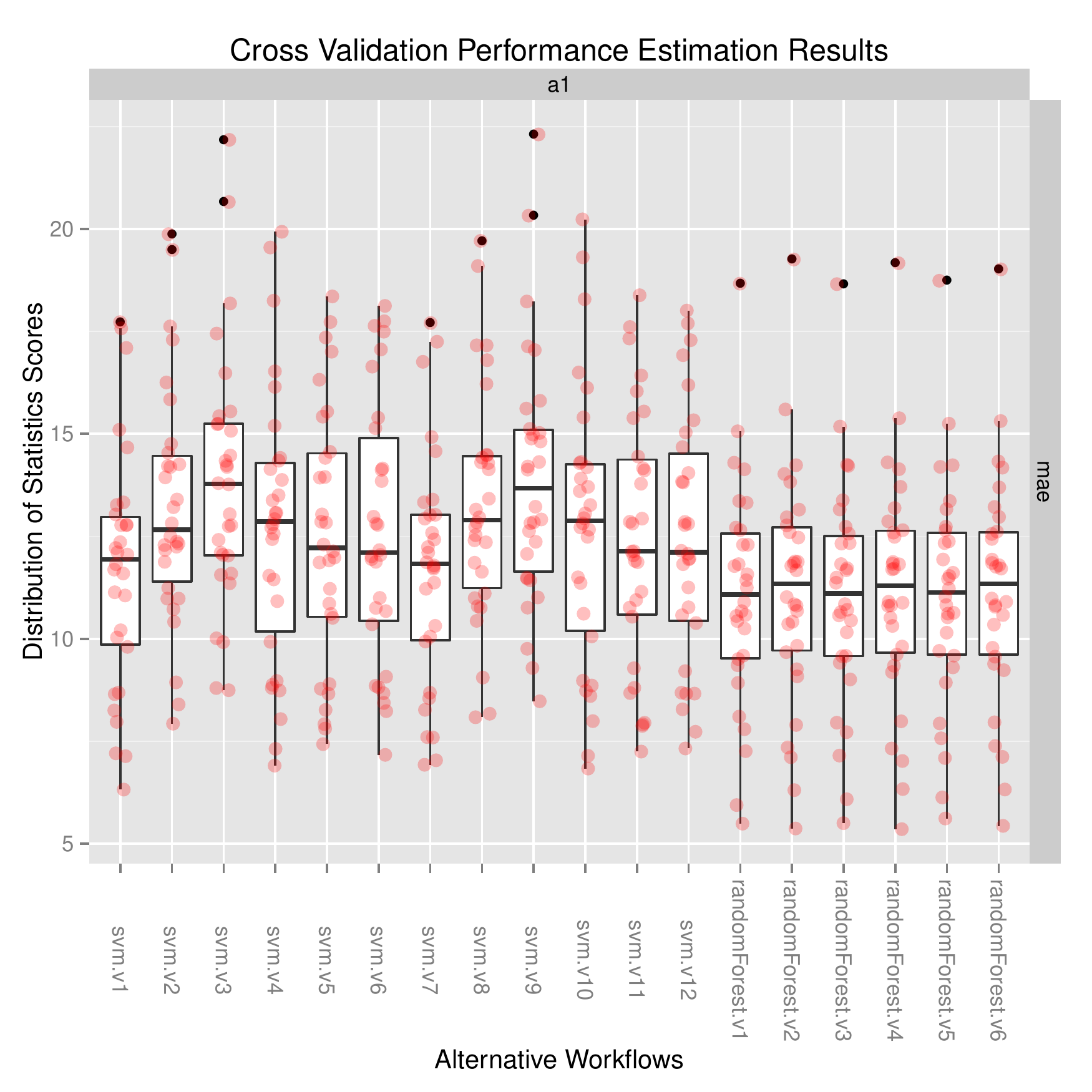} 

}

\caption[The MAE results for the task ``a1'']{The MAE results for the task ``a1''.\label{fig:maeA1}}
\end{figure}

\end{knitrout}
 
As before we are using the generic function \texttt{plot()} but this
time applied to a subset of the original object with all results. This
subset is obtained using the generic function \texttt{subset()} that
accepts several parameters to specify the subset we are interested
on. In this case we are using the parameters \texttt{tasks} and
\texttt{metrics} to indicate that we want to analyze only the results
concerning the task ``a1'' and the metric ``mae''. Other possibility
is the parameter \texttt{workflows} for indicating a subset of the
workflows. Both \texttt{workflows}, \texttt{tasks} and \texttt{metrics} accept
as values a character string containing a regular expression that will
be used internally with the R function \texttt{grep()} over the vector
of names of the respective objects (names of the workflows, names of
the tasks and names of the metrics, respectively). For instance, if
you want to constrain the previous graph even further to the workflows
whose name ends in ``4'' (absurd example of course!), you could use
the following:

\begin{knitrout}\footnotesize
\definecolor{shadecolor}{rgb}{0.969, 0.969, 0.969}\color{fgcolor}\begin{kframe}
\begin{alltt}
\hlkwd{plot}\hlstd{(}\hlkwd{subset}\hlstd{(res,} \hlkwc{tasks}\hlstd{=}\hlstr{'a1'}\hlstd{,} \hlkwc{workflows}\hlstd{=}\hlstr{'4$'}\hlstd{))}
\end{alltt}
\end{kframe}\begin{figure}[]

{\centering \includegraphics[width=0.75\textwidth]{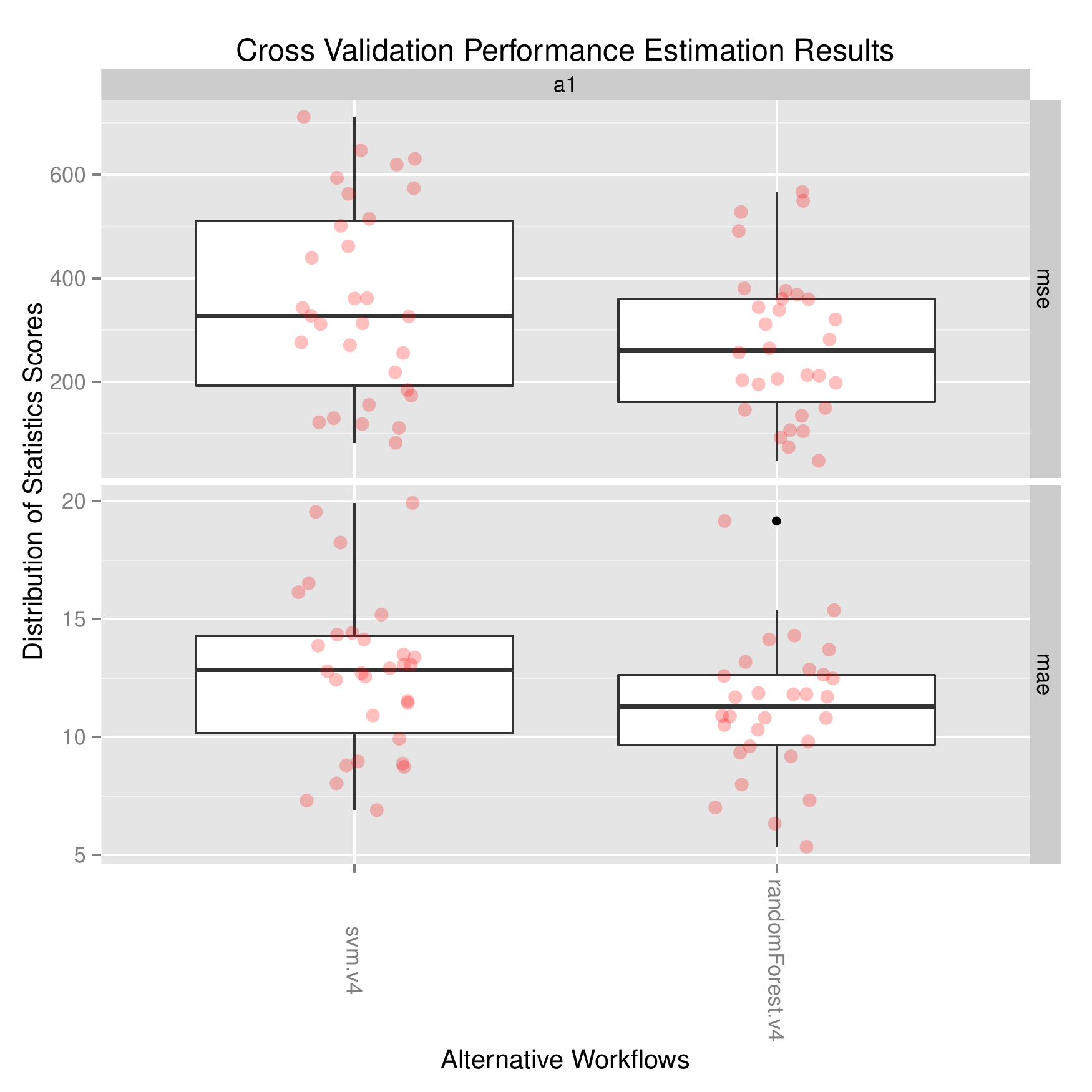} 

}

\caption[Illustration of the use of regular expressions in sub-setting the results objects]{Illustration of the use of regular expressions in sub-setting the results objects.\label{fig:mseA1b}}
\end{figure}

\end{knitrout}

If you are more familiar with the syntax of "wildcards" you may use
the R function \texttt{glob2rx()} to convert to regular expressions,
as show in the following example:

\begin{knitrout}\small
\definecolor{shadecolor}{rgb}{0.969, 0.969, 0.969}\color{fgcolor}\begin{kframe}
\begin{alltt}
\hlkwd{summary}\hlstd{(}\hlkwd{subset}\hlstd{(res,} \hlkwc{tasks}\hlstd{=}\hlstr{'a1'}\hlstd{,} \hlkwc{workflows}\hlstd{=}\hlkwd{glob2rx}\hlstd{(}\hlstr{'*svm*'}\hlstd{),}\hlkwc{metrics}\hlstd{=}\hlstr{'mse'}\hlstd{))}
\end{alltt}
\begin{verbatim}
## 
## == Summary of a  Cross Validation Performance Estimation Experiment ==
## 
## Task for estimating  mse  using
##  3 x 10 - Fold Cross Validation
## 	 Run with seed =  1234 
## 
## * Predictive Tasks ::  a1
## * Workflows  ::  svm.v1, svm.v2, svm.v3, svm.v4, svm.v5, svm.v6, svm.v7, svm.v8, svm.v9, svm.v10, svm.v11, svm.v12 
## 
## -> Task:  a1
##   *Workflow: svm.v1 
##               mse
## avg     305.37037
## std     147.62784
## med     275.08382
## iqr     211.22466
## min      79.09407
## max     605.36798
## invalid   0.00000
## 
##   *Workflow: svm.v2 
##              mse
## avg     349.1888
## std     171.6171
## med     301.5451
## iqr     196.5386
## min     120.9873
## max     821.4265
## invalid   0.0000
## 
##   *Workflow: svm.v3 
##              mse
## avg     367.1953
## std     180.4110
## med     345.6525
## iqr     203.3054
## min     142.0827
## max     915.9872
## invalid   0.0000
## 
##   *Workflow: svm.v4 
##               mse
## avg     356.54033
## std     187.29360
## med     326.86720
## iqr     318.83908
## min      82.03737
## max     711.67485
## invalid   0.00000
## 
##   *Workflow: svm.v5 
##              mse
## avg     376.3833
## std     235.3738
## med     282.9879
## iqr     308.7286
## min     104.9834
## max     975.8439
## invalid   0.0000
## 
##   *Workflow: svm.v6 
##                mse
## avg      396.15700
## std      287.60697
## med      285.41018
## iqr      294.37250
## min       98.40015
## max     1260.52110
## invalid    0.00000
## 
##   *Workflow: svm.v7 
##               mse
## avg     304.23863
## std     144.75696
## med     271.52274
## iqr     212.00615
## min      78.69374
## max     587.86503
## invalid   0.00000
## 
##   *Workflow: svm.v8 
##              mse
## avg     355.5680
## std     176.4419
## med     313.3674
## iqr     190.9779
## min     130.2359
## max     862.8762
## invalid   0.0000
## 
##   *Workflow: svm.v9 
##              mse
## avg     375.2226
## std     188.0744
## med     350.1804
## iqr     195.8608
## min     143.0326
## max     954.8950
## invalid   0.0000
## 
##   *Workflow: svm.v10 
##               mse
## avg     359.45788
## std     188.57821
## med     331.28706
## iqr     320.80755
## min      82.84897
## max     721.55851
## invalid   0.00000
## 
##   *Workflow: svm.v11 
##              mse
## avg     367.1249
## std     216.2313
## med     282.4140
## iqr     288.6100
## min     101.2520
## max     959.5912
## invalid   0.0000
## 
##   *Workflow: svm.v12 
##                mse
## avg      376.05576
## std      244.87929
## med      284.02493
## iqr      278.97164
## min       94.36646
## max     1179.99337
## invalid    0.00000
\end{verbatim}
\end{kframe}
\end{knitrout}

The following are some illustrations of the use of other available
utility functions.

Obtaining the scores on all iterations and metrics of a workflow on a
particular task:

\begin{knitrout}\footnotesize
\definecolor{shadecolor}{rgb}{0.969, 0.969, 0.969}\color{fgcolor}\begin{kframe}
\begin{alltt}
\hlkwd{getScores}\hlstd{(res,} \hlstr{'svm.v6'}\hlstd{,}\hlstr{'a3'}\hlstd{)}
\end{alltt}
\begin{verbatim}
##              mse      mae
##  [1,]  58.773058 4.039651
##  [2,]  63.420588 4.836262
##  [3,] 124.212723 6.629852
##  [4,]  20.818437 2.902345
##  [5,]  28.507875 3.202152
##  [6,]  18.424942 2.774288
##  [7,]  14.446194 2.509190
##  [8,]  54.975336 4.584341
##  [9,]  23.521575 2.871156
## [10,] 135.025071 6.735133
## [11,] 112.234104 5.328625
## [12,]  81.117165 5.569711
## [13,]  34.504072 3.779135
## [14,]   9.790665 2.162704
## [15,]  30.126166 3.476657
## [16,]  58.715226 4.654773
## [17,]  23.793327 3.171367
## [18,]  26.480963 3.236722
## [19,]  99.572929 5.928280
## [20,]  27.569010 3.293526
## [21,]   8.495628 2.129290
## [22,]  27.933235 3.197012
## [23,]  20.164491 2.850632
## [24,]  72.373337 5.364697
## [25,]  69.470293 4.813790
## [26,] 126.075926 5.477111
## [27,]  65.167023 5.459279
## [28,]   9.437532 2.299811
## [29,]  22.696736 3.511973
## [30,]  90.988401 5.114133
\end{verbatim}
\end{kframe}
\end{knitrout}

Getting the summary of the results of a particular workflow on a  predictive task :

\begin{knitrout}\footnotesize
\definecolor{shadecolor}{rgb}{0.969, 0.969, 0.969}\color{fgcolor}\begin{kframe}
\begin{alltt}
\hlkwd{estimationSummary}\hlstd{(res,}\hlstr{'svm.v3'}\hlstd{,} \hlstr{'a7'}\hlstd{)}
\end{alltt}
\begin{verbatim}
##                mse      mae
## avg      30.516360 3.156927
## std      29.265924 1.213391
## med      23.096086 3.016725
## iqr      31.305352 1.053218
## min       3.083724 1.393694
## max     141.576530 7.615879
## invalid   0.000000 0.000000
\end{verbatim}
\end{kframe}
\end{knitrout}

Finally, the \texttt{metricsSummary()} function allows you to apply any
summary function (defaulting to \texttt{mean()}) to the iterations estimates. The following
calculates the median of the results of the SVMs on the task ``a1'',

\begin{knitrout}\small
\definecolor{shadecolor}{rgb}{0.969, 0.969, 0.969}\color{fgcolor}\begin{kframe}
\begin{alltt}
\hlkwd{metricsSummary}\hlstd{(}\hlkwd{subset}\hlstd{(res,} \hlkwc{workflows}\hlstd{=}\hlkwd{glob2rx}\hlstd{(}\hlstr{'*svm*'}\hlstd{),} \hlkwc{tasks}\hlstd{=}\hlstr{'a1'}\hlstd{),}
               \hlkwc{summary}\hlstd{=}\hlstr{'median'}\hlstd{)}
\end{alltt}
\begin{verbatim}
## $a1
##        svm.v1    svm.v2    svm.v3    svm.v4    svm.v5    svm.v6    svm.v7    svm.v8
## mse 275.08382 301.54507 345.65254 326.86720 282.98787 285.41018 271.52274 313.36744
## mae  11.93685  12.65768  13.78127  12.85777  12.21762  12.10388  11.82759  12.89971
##        svm.v9   svm.v10   svm.v11   svm.v12
## mse 350.18044 331.28706 282.41397 284.02493
## mae  13.67427  12.87902  12.13263  12.10956
\end{verbatim}
\end{kframe}
\end{knitrout}

\section{Conclusions}

We have presented \PE that aims at being a general package for estimating and comparing the performance of any approach to any predictive task in R. The package allows users to very easily carry out standard out-of-the-box comparative experiments between existing modeling tools in R. However, it also allows the users to supply their own functions implementing special workflows to solve tasks. This means that the package should cover the needs of occasional users as well as advanced users that which to try and compared their own proposed workflows.

The \PE also includes several facilities for testing the statistical significance of the observed differences, namely implementing the current state of the art in this subject as described in~\cite{Dem06}, including CD diagrams for both the Nemenyi and Bonferroni-Dunn post-hoc tests.

Finally, we have provided a few illustrations of how to use \PE for larger experiments, namely taking advantage of parallel computation that is available in R.

\bibliographystyle{alpha}
\bibliography{performanceEstimation}
\end{document}